\documentclass[journal]{IEEEtran}

\usepackage{amsmath,amsfonts,bm}
\usepackage{array}
\usepackage{textcomp}
\usepackage{stfloats}
\usepackage{url}
\usepackage{verbatim}
\usepackage{graphicx}
\usepackage{cite}
\usepackage{amssymb}
\usepackage[ruled,linesnumbered, noend]{algorithm2e}
\usepackage{booktabs}
\usepackage{diagbox}

\ifCLASSOPTIONcompsoc
\usepackage[caption=false, font=footnotesize, labelfont=sf, textfont=sf,subrefformat=parens]{subfig}
\else
\usepackage[caption=false, font=footnotesize,,subrefformat=parens]{subfig}
\usepackage{makecell}
\usepackage{multirow}
\usepackage{xcolor}
\usepackage{tabularx}
\usepackage{bbding}
\usepackage{threeparttable}

\usepackage{ltablex} 
\usepackage{lipsum}
\usepackage{ragged2e}
\usepackage{pifont}     
\usepackage{array}      

\begin{document}

\title{Multi-objective Low-altitude IRS-assisted ISAC Optimization via Generative AI-enhanced Deep Reinforcement Learning}

\author{
        Wenwen Xie,
        Geng Sun,~\IEEEmembership{Senior Member,~IEEE},
        Chuang Zhang,
        Hongyang Du, \\
        Kaibin Huang,~\IEEEmembership{Fellow,~IEEE}, 
        Victor C. M. Leung,~\IEEEmembership{Life Fellow,~IEEE}
        \thanks
        {
        \par Wenwen~Xie, Geng~Sun, and Chuang~Zhang are with the College of Computer Science and Technology, Jilin University, Changchun 130012, China~(e-mail: xieww22@mails.jlu.edu.cn, sungeng@jlu.edu.cn, chuangzhang1999@gmail.com).
        \par Hongyang Du and Kainbin Huang are with the Department of Electrical and Electronic Engineering, The University of Hong Kong, Hong Kong 999077, China~(emails: duhy@eee.hku.hk, huangkb@hku.hk).
        \par Victor C.M. Leung is with the Artificial Intelligence Research Institute, Shenzhen MSU-BIT University, Shenzhen 518115, China, with the College of Computer Science and Software Engineering, Shenzhen University, Shenzhen 518060, China, and also with the Department of Electrical and Computer Engineering, The University of British Columbia, Vancouver V6T 1Z4, Canada (e-mail: leung@ieee.org).
       \par (\textit{Corresponding author: Geng Sun.})
        }
 }

\maketitle

\begin{abstract}
    \par Integrated sensing and communication (ISAC) has attracted increasing attention due to its critical importance in enabling sixth-generation (6G) wireless networks. Nonetheless, simultaneously optimizing communication and sensing performance within a unified ISAC framework poses an inherent challenge. In this work, we investigate a low-altitude intelligent reflecting surface (IRS)-assisted ISAC architecture, in which a base station (BS) performs dual functions by serving multiple communication users while sensing an obstructed target, with link quality enhanced via an IRS deployed on an uncrewed aerial vehicle (UAV). Moreover, we formulate a multi-objective optimization problem, which aims to maximize the communication rate of the users and the beampattern gain of the target while minimizing UAV propulsion energy consumption by jointly optimizing the BS beamforming matrix, IRS phase shifts, UAV flight velocity, and UAV flight angle. Considering the non-convexity, trade-off, and long-term characteristics of the formulated optimization problem, we propose a generative diffusion model-based deep deterministic policy gradient (GDMDDPG) algorithm to solve the problem. Specifically, the diffusion model is incorporated into the actor network of the deep deterministic policy gradient (DDPG) to improve the action quality, with a noise perturbation mechanism for better exploration and a recent prioritized experience replay (RPER) sampling mechanism for enhanced training efficiency. The simulation results indicate that the GDMDDPG algorithm delivers superior performance compared to existing methods.
\end{abstract}


\section{Introduction}
\label{sec:Introduction}
    
\par As fifth-generation (5G) networks are commercially deployed, and research on sixth-generation (6G) progresses, conventional communication systems have revealed notable shortcomings in meeting the increasingly diverse and growing demands~\cite{Yang2026}. In particular, as the low-altitude economy continues to evolve~\cite{10759668}, leveraging airspace for activities such as logistics and rescue operations, the need for seamless communication and real-time sensing becomes even more critical. Therefore, it is necessary to adopt a new approach that moves beyond traditional communication-centric network designs and embraces integrated sensing-communication networks. In this case, integrated sensing and communication (ISAC) technology stands out as a practical and efficient solution~\cite{Qaisar2026}, and it has been recognized by the International Telecommunication Union (ITU) as a critical application scenario among the six primary use cases envisioned for the 6G era~\cite{Kaushik2024}.

\par In the practical implementation of ISAC systems, non-ideal propagation environments caused by obstacle occlusion can significantly affect system performance~\cite{Geng2025}. In this case, the intelligent reflecting surface (IRS) emerges as a promising solution~\cite{Chen2026}. Specifically, IRS consists of a planar array with densely packed sub-wavelength reflective elements, each of which can individually modify the phase and amplitude of incoming signals~\cite{Xie2026}. Because each reflective element is small in size, an IRS with a practical size can incorporate numerous elements to enable substantial beamforming gains via co-modulation, effectively compensating for path loss~\cite{Hua2024}. Thus, the unique capability for remodeling the propagation environment provides an innovative solution to enhance the robustness of the ISAC system in occlusion scenarios~\cite{Sankar2024}. Moreover, forming a low-altitude IRS by integrating the IRS in a maneuverable unrewed aerial vehicle (UAV) can overcome the spatial coverage limitation of the traditional fixed IRS, further enhancing the effectiveness of the ISAC system~\cite{Xu2024}.

\par Currently, the ISAC system can mainly be classified into two major architectures, \textit{i.e.}, radar-communication coexistence (RCC) architecture and dual-functional radar-communication (DFRC) architecture. In the former, radar transceivers and communication transmitters are physically distributed across distinct locations~\cite{He2022}, which typically leads to severe co-channel interference issues. Moreover, to achieve effective coordination among radar and communication functionalities, the RCC architecture requires complex information feedback mechanisms, inevitably introducing substantial communication overhead. In contrast, the DFRC architecture enables simultaneous communication and sensing functionalities by sharing the same hardware resources, thereby improving resource utilization~\cite{Liu2022}.

\par However, designing an effective beamforming and resource allocation strategy addressing the competing demands between the communication and sensing functionalities in DRFC architecture remains challenging~\cite{Zuo2023}. In particular, in the low-altitude IRS-assisted ISAC system, dynamic characteristics are introduced to both the application scenario and optimization problem due to the time-varying nature of channels and the mobility of the low-altitude IRS. However, obtaining accurate prior knowledge in advance in the dynamic scenario is difficult, which makes traditional optimization methods (\textit{e.g.}, convex optimization) show limitations in solving such dynamic optimization problems~\cite{Guo2023}. In this case, deep reinforcement learning (DRL) provides a promising solution. Specifically, DRL enables the agent to learn by interacting with the environment, removing the necessity for prior information and attaining performance close to optimal~\cite{Sun2025a}. However, traditional DRL algorithms employ Multi-Layer Perceptron (MLP)-based actor networks, which struggle to capture the complex environment characteristics of the low-altitude IRS-assisted ISAC system~\cite{sun2025-1}. Moreover, DRL algorithms typically necessitate a substantial number of agent-environment interactions to achieve meaningful learning progress.

\par To address the above challenges, we propose a novel generative artificial intelligence (GenAI)-enabled DRL algorithm. To the best of our knowledge, no existing work simultaneously optimizes communication, sensing, and energy efficiency performance in the movable low-altitude IRS-assisted ISAC system. The main contributions of this paper are outlined as follows:

\begin{itemize}
    \item \textbf{Low-altitude IRS-assisted ISAC System}: A low-altitude IRS-assisted ISAC system is explored, where a dual-functional base station (BS) equipped with multiple antennas simultaneously communicates with multiple users and provides sensing services to an occluded target. Moreover, to mitigate the impact of obstacles on the sensing performance of the BS, the IRS is mounted on a UAV to form a low-altitude IRS. Compared with fixed IRS installations, low-altitude IRS offers greater spatial flexibility, enabling the dynamic establishment of virtual line-of-sight (LoS) links in response to environment changes, thereby significantly improving the channel quality of the system.
    
    \item \textbf{Formulation of Multi-objective Optimization Problem}: We formulate a multi-objective optimization problem to maximize the communication rate of users and the beampattern gain of the target while minimizing the UAV propulsion energy consumption through joint optimization of the BS active beamforming matrix, IRS phase shifts, UAV flight velocity, and UAV flight angle. Notably, the formulated optimization problem is a non-convex dynamic optimization problem, which limits traditional optimization algorithms in solving the problem.
    \item \textbf{GenAI-enabled DRL Algorithm}: We first transform the formulated optimization problem into a Markov decision process (MDP), and then propose a generative diffusion model-based deep deterministic policy gradient (GDMDDPG) algorithm, which is a GenAI-enabled DRL algorithm to solve the problem. Specifically, the diffusion model-based actor network integrates the analytical and generative capabilities of the diffusion model, enhancing both environment state analysis and action generation of the algorithm. Moreover, the noise perturbation mechanism introduces stochasticity to overcome policy stagnation and foster exploration within the action space. Meanwhile, the recent prioritized experience replay (RPER) mechanism combines recent experience emphasis (ERE) and prioritized experience replay (PER) to dynamically select high-value recent experiences during network updates, which accelerates the learning process of the algorithm.
    \item \textbf{Simulation and Analysis}: Simulation results highlight the advantages of the proposed GenAI-enabled DRL approach. In particular, it significantly outperforms three deployment strategies and two representative DRL baselines across diverse simulation scenarios, achieving enhanced communication and sensing performance with lower UAV energy consumption. Moreover, we perform a simulation analysis of the proposed algorithm across various initial UAV locations to further validate its robustness.
\end{itemize}

\par The organization of this paper is as follows: Section~\ref{sec: Related Work} introduces the related work. Section~\ref{sec: System Model and Problem Formulation} presents the system model. Section~\ref{sec: Problem Analyses} formulates and analyzes the optimization problem. Section~\ref{section: Proposed Algorithm} provides the proposed GenAI-enabled DRL algorithm. Section~\ref{sec: Simulations} conducts the simulations. Finally, the paper is summarized in Section~\ref{sec: Conclusion}.

%
%
\section{Related Work}
\label{sec: Related Work}

\par In this section, we present the existing work related to ISAC from three perspectives, \textit{i.e.}, scenario, performance metrics, and optimization methods.

\subsection{IRS-assisted ISAC System}
\par As the demand for deep integration of communication and sensing continues to grow in the 6G era, ISAC has emerged as a key research topic. For example, the authors in~\cite{Hua2023} studied a classic ISAC system, where the beamforming matrix of the dual-functional BS is optimized to improve the quality of communication and sensing services. A mobile antenna-enabled ISAC system is studied in~\cite{Lyu2025}, which utilizes flexible antenna layout capabilities to improve the beamforming accuracy of the BS, thereby improving system performance in the clutter environment. However, the aforementioned works ignored the negative effect of obstacles on ISAC performance, which may make them unsuitable for densely built-up scenarios such as urban areas. In this case, flexible low-altitude platforms such as UAVs have become a widely recognized solution with the rise of the low-altitude economy. For example, the authors in~\cite{Ning2026,11543384} provided forward-looking insights and feasibility analyses on the UAV acting as the BS in urban environments to perform ISAC tasks, including accident detection, delivery, and pedestrian awareness.

\par Given the powerful capabilities of the IRS in reconstructing the wireless propagation environment, the IRS is introduced to ISAC systems for improving the channel quality. For example, a distributed semi-passive IRS-assisted single-input multiple-output (SIMO)-based ISAC system is investigated~\cite{Hu2023}, where one sub-IRS supports data uploading from a blocked single-antenna user to the BS and the other two sub-IRS perform user positioning in the sensing mode. The authors in~\cite{Li2023} deployed an active IRS equipped with amplifiers to assist a multiple-input multiple-output (MIMO)-based ISAC scenario. Moreover, the authors in~\cite{Liao2023} studied an ISAC scheme in which the BS senses multiple blocked targets in the clutter environment with the assistance of IRS, and the results indicate the effectiveness of IRS in enhancing the reliability of blocked target sensing. In addition, a multi-IRS-assisted ISAC system was studied in~\cite{Fang2024}, and the authors conducted independent analyses of the system performance in two blocked sensing scenarios involving point targets and extended targets, respectively.

\par However, some of the aforementioned existing works deployed mobile aerial BS to mitigate obstacle effects and perform ISAC tasks, which may limit further performance improvements in resource-intensive applications due to the resource constraints of the aerial platform. In this case, most of the existing works utilized static IRSs to reconstruct the wireless propagation environment to improve the ISAC performance in obstructed environments. Notably, deploying flexible and mobile IRSs can further unleash the performance of IRS-assisted ISAC systems in complex wireless environments by enabling more flexible signal propagation links.

\subsection{Performance Metrics in IRS-assisted ISAC System}

\par Different performance metrics in the ISAC system demonstrate different optimization focuses. In terms of communication metrics, the authors in~\cite{Wang2023,Xu2024} optimized resource allocation for the IRS-supported ISAC system to improve the communication rate for users. Moreover, the authors in~\cite{11395974,Jiang2024} tackled eavesdropping threats in the ISAC scenario by designing the BS beamforming matrix and adjusting the IRS phase shifts to enhance the secure communication rate. For sensing metrics, the authors in~\cite{Zuo2023} explored serving users and targets simultaneously by transmitting superimposed NOMA signals, and aimed to maximize the beampattern gain to improve the energy concentration on the target. Moreover, the authors in~\cite{Long2024} tackled the vehicle positioning and high-quality communication challenges by optimizing IRS phase shifts to minimize the position error bound. For energy metrics, the authors in~\cite{11159297} aimed to reduce the overall energy consumption of a UAV formations-enabled ISAC system by minimizing the maximum linear quadratic regulator (LQR).

\par Moreover, some existing works considered optimizing multiple performance metrics to jointly improve the communication and sensing capabilities of the ISAC system. For example, the authors in~\cite{11165100} optimized the total energy consumption, communication rate, and computation rate to improve the energy efficiency and computation capability of the IRS-assisted ISAC system. Moreover, a multi-UAV-assisted ISAC framework was studied in~\cite{Wu2023}, in which UAV trajectories and scheduling policies are jointly designed to maximize the downlink data throughput while reducing the Cramér-Rao bound. In addition, the authors in~\cite{Jin2025} utilized the UAVs as dual-functional BSs to perform ISAC tasks, and minimized the minimum LQR cost while maximizing the Fisher information matrix to reduce the energy consumption and improve the sensing accuracy of the system.

\par However, few studies simultaneously optimize communication, sensing, and energy metrics in UAV-enabled ISAC systems, and such joint optimization can maintain sensing and communication capabilities while significantly improving energy efficiency.

\subsection{Optimization Methods in IRS-assisted ISAC System}

\par A variety of optimization methods have been extensively applied in IRS-assisted ISAC scenarios to address the associated design challenges. For example, a multi-strategy alternating optimization algorithm based on quadratically constrained quadratic programming and semidefinite relaxation (SDR) was adopted in~\cite{Hu2024} to enhance both the transmission rate and probing power. Moreover, the authors in~\cite{11177504} adopted the alternative optimization method that integrates the successive convex approximation and SDR to improve the secrecy rate of the secure IRS-assisted ISAC system. Moreover, the authors in~\cite{Cao2023} designed an improved particle swarm optimization algorithm to jointly optimize active and passive beamforming for maximizing the echo signal power. 

\par Since DRL is well-suited for solving dynamic sequential optimization problems, it has been widely applied to decision making in IRS-assisted ISAC systems. For example, the authors in~\cite{10594249} employed the proximal policy optimization (PPO) algorithm to optimize the beamforming matrix and the UAV trajectory to maximize the transmission rate in the low-altitude IRS-assisted ISAC system. Furthermore, the authors in~\cite{10257639} explored a STAR-RIS-supported secure ISAC framework and used the soft actor-critic (SAC) and DDPG algorithms to improve the long-term secrecy rate.

\par However, convex optimization and swarm intelligence approaches encounter challenges in dynamic environments where acquiring prior knowledge is difficult~\cite{Guo2023}. Moreover, although DRL algorithms have inherent advantages over conventional optimization methods, the standard DRL algorithms above still have limitations in analyzing complex environmental features and state-action relationships, which may affect the quality of decision-making.

%
\section{System Model}
\label{sec: System Model and Problem Formulation}

\par In this section, we provide detailed models of communication, sensing, and UAV mobility in the low-altitude IRS-assisted ISAC system.

\subsection{System Overview}
\par
\begin{figure}[t]
    \centering
    \includegraphics[width=0.95\linewidth]{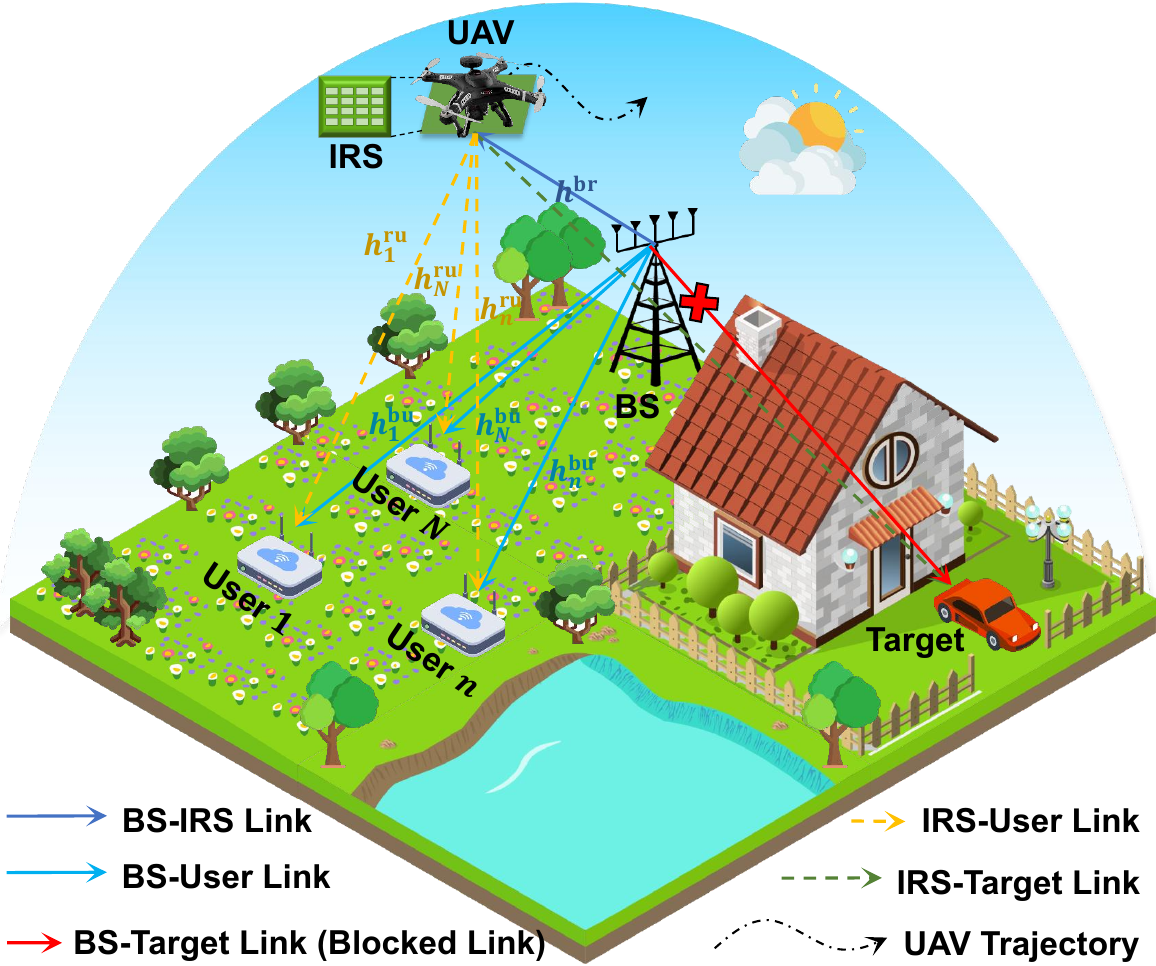}
    \caption{Low-altitude IRS-assisted ISAC system. In this system, the BS provides communication services to multiple users and performs sensing tasks for a target with the assistance of a low-altitude IRS.}
    \label{fig: system model}
\end{figure}

 As shown in Fig.~\ref{fig: system model}, a low-altitude IRS-assisted ISAC system is considered, which consists of a BS equipped with a uniform antenna array of $M$ antennas, a ground user set denoted as $\mathcal{N} = \{1, \ldots, N\} $, and a ground target. Specifically, the dual-functional BS communicates with $N$ users while performing sensing tasks for the target. It is assumed that obstacles block the LoS path from the BS to the target, which substantially degrades sensing performance at the BS~\cite{Xu2024}. In this case, we deploy a UAV-carried IRS as a low-altitude IRS to establish virtual LoS links between the BS and the target, which can mitigate the blocking effect of existing obstacles. The altitude IRS consists of $L=L_{\text{x}} \times L_{\text{y}}$ reflective elements, where $L_{\text{x}}$ and $L_{\text{y}}$ represent the number of reflective elements in each row and column, respectively. As such, the low-altitude IRS is capable of enhancing both sensing and communication performance by improving signal quality, which boosts a more flexible and robust system. 

\par We discretize the operation time of the system, dividing the total time $T_{\text{d}}$ into $T$ equal-length time slots, denoted as $\mathcal{T}=\{1, \ldots, T\}$, with the length of each time slot $t_{\text{d}}=\frac{T_{\text{d}}}{T}$. Moreover, the widely adopted Cartesian coordinate system is used to depict the spatial locations of all elements in our considered scenario. Specifically, due to the maneuverability of the UAV, the location of the low-altitude IRS is variable, and it can be represented as $\bm{q}^{\text{r}}[t]=(x^{\text{r}}[t], y^{\text{r}}[t], z^{\text{r}})$ in time slot $t$. Similarly, the locations of the BS, $n$-th user and the target are denoted as $\bm{q}^{\text{b}}=(x^{\text{b}}, y^{\text{b}}, z^{b})$, $q_{n}^{\text{u}}=(x_{n}^{\text{u}}, y_{n}^{\text{u}}, 0)$, $\bm{q}^{\text{g}}=(x^{\text{g}}, y^{\text{g}}, 0 )$, respectively.

\par Notably, multiple sensing and communication links change as the UAV trajectory varies, which introduces high dynamics to the considered system. In the following, we provide a detailed model for sensing, communication, and UAV mobility to capture the key optimization variables and objectives.

\subsection{Channel Model}
\par The channel between the BS and the user is comprised of two main components, \textit{i.e.}, the direct link and the reflected link assisted by the low-altitude IRS. Next, a detailed description for the channel model is provided.

\subsubsection{Direct Link}
\par Due to the intricate nature of signal propagation in the considered scenario, we utilize the Rician model to represent direct links, effectively capturing both the LoS component and scattering effects from multipath propagation~\cite{Xu2024,AlHourani2014}. Then, the channel vector $\bm{h}_{n}^{\text{bu}}[t] \in \mathbb{C}^{M \times 1}$ from the BS to $n$-th user in time slot $t$ can be represented as
\begin{equation}
    \label{equ_1}
    \bm{h}_{n}^{\text{bu}}[t] = \sqrt{\frac{L_{0}}{(d_{n}^{\text{bu}})^{\iota^{\text{bu}}}}}  \big(\sqrt{\frac{\eta^\text{{bu}}}{\eta^{\text{bu}}+1}}\bm{g}_{n,\text{L}}^{\text{bu}}[t] + \sqrt{\frac{1}{\eta^{\text{bu}}+1}}\bm{g}_{n,\text{N}}^{\text{bu}}[t] \big),
\end{equation}

\noindent where $L_{0}$ represents the channel power gain at a reference distance of 1 m, $\iota^{\text{bu}}$ denotes the path loss, and $\eta^{\text{bu}}$ is the Rician factor. Moreover, $d_{n}^{\text{bu}}=\sqrt{\|q^{\text{b}} - q_{n}^{\text{u}} \|^{2}}$ represents the distance between the BS and the $n$-th user. In addition, $\bm{g}_{n,\text{L}}^{\text{bu}}[t]$ denotes the LoS component, and $\bm{g}_{n, \text{N}}^{\text{bu}}[t] \sim \mathcal{CN}(0, \mathbf{I}_{M \times 1})$ is the none-line-of-sight (NLoS) component that follows the complex Gaussian distribution, and $\mathbf{I}_{M \times 1}$ is the identity matrix.

\subsubsection{Reflected Link}
\par Note that the reflected link consists of the BS-low-altitude IRS link and the low-altitude IRS-users link, respectively.
\par When the low-altitude IRS is deployed at a sufficiently high altitude, the air-ground links are largely governed by the LoS component~\cite{AlHourani2014}. Then, the channel vector $\bm{H}^{\text{br}}[t] \in \mathbb{C}^{L \times M}$ between the BS and the low-altitude IRS in time slot $t$ is given by
\begin{equation}
    \label{equ_2}
    \bm{H}^{\text{br}}[t] = \sqrt{\frac{L_{0}}{(d^{\text{br}}[t])^{\iota^{\text{br}}}}} \bm{\alpha}_{\text{br}}[t]\bm{\alpha}_{\text{b}}^{H}[t],
\end{equation}
\noindent where $\iota^{\text{br}}$ is the path loss in the BS-IRS link, and $d^{\text{br}}[t]=\sqrt{\|q^{\text{b}}-q^{\text{r}}[t]\|^{2}}$ represents the distance between the BS and the low-altitude IRS in time slot $t$. Moreover, $\bm{\alpha}_{\text{b}}[t]$ and $\bm{\alpha}_{\text{br}}[t]$ denote the response vectors of the BS and low-altitude IRS in time slot $t$, respectively. Moreover, $(\cdot)^{H}$ denotes the Hermitian transpose operations, respectively.

\par Likewise, the channel vector $\bm{h}_{n}^{\text{ru}}[t] \in \mathbb{C}^{L \times 1}$ represents the communication link between the low-altitude IRS and the $n$-th user in time slot $t$, which is given by
\begin{equation}
    \label{equ_4}
    \bm{h}_{n}^{\text{ru}}[t] = \sqrt{\frac{L_{0}}{(d_{n}^{\text{ru}}[t])^{\iota^{\text{ru}}}}} \bm{\alpha}_{n,\text{L}}^{\text{ru}}[t],
\end{equation}

\noindent where $\alpha^{\text{ru}}$ indicates the path loss. Moreover, $d_{n}^{\text{ru}}[t]=\sqrt{\|q^{\text{r}}[t]-q_{n}^{\text{u}}\|^{2}}$ is used to represent the distance from the low-altitude IRS and the $n$-th user. Moreover, $\bm{\alpha}_{n,\text{L}}^{\text{ru}}[t]$ denotes the LoS component in time slot $t$.

\subsubsection{Composite Link}
\par The diagonal phase-shift matrix of the IRS in time slot $t$ can be represented as
\begin{equation}
    \label{equ_6}
    \bm{\Phi}[t] = \rm{diag}\left(\bm{\widetilde{\theta}}[t]\right),
\end{equation}

\noindent where $\bm{\widetilde{\theta}}[t]=\left[e^{j\theta_{1,1}[t]}, \ldots, e^{j\theta_{l_x,l_y}[t]}, \ldots, e^{j\theta_{L_x,L_y}[t]}\right]^{T}$ is the passive beamforming vector of the IRS, and $\theta_{l_x,l_y}[t] \in [0, 2\pi)$ is the phase shift of the $(l_x,l_y)$-th IRS element, with $l_x \in \mathcal{L}_x$ and $l_y \in \mathcal{L}_y$, where $\mathcal{L}_x = \{1,2,\ldots,L_x\}$ and $\mathcal{L}_y = \{1,2,\ldots,L_y\}$.

\par As such, the composite channel between the BS and the $n$-th user is given by
\begin{equation}
    \label{equ_7}
    \widetilde{\bm{h}}_{n}^{\text{bu}}[t]=\left(\bm{h}_{n}^{\text{bu}}[t]\right)^{H} + \left(\bm{h}_{n}^{\text{ru}}[t]\right)^{H}\bm{\Phi}[t]\bm{H}^{\text{br}}[t].
\end{equation}

\subsection{Communication Model}
\par Let $s_{n}[t]$ and $\bm{\omega}_{n}[t] \in \mathbb{C}^{M \times 1}$ denote the symbol transmitted by the BS to the $n$-th user and the corresponding active beamforming vector in time slot $t$, respectively. We consider that the symbols are independent and identically distributed with zero mean and unit variance, expressed as $s_{n} \sim \mathcal{N}(0, 1)$. Therefore, the BS transmit signal can be represented by
\begin{equation}
    \label{equ_8}
    \mathbf{x}[t] = \sum_{n=1}^{N}\bm{\omega}_{n}[t]s_{n}[t].
\end{equation}

\par Then, the signal received by the $n$-th user in time slot $t$ is given by
\begin{eqnarray}
    \label{equ_9}
    y_{n}[t] = \underbrace{\widetilde{\bm{h}}_{n}^{\text{bu}}[t]\bm{\omega}_{n}[t]s_{n}[t]}_{\text{desired \ signal}} + \underbrace{\sum\nolimits_{i \neq n}^{N} \widetilde{\bm{h}}_{n}^{\text{bu}}[t]\bm{\omega}_{i}[t]s_{i}[t]}_{\text{interference}} + \underbrace{\varrho_{n}[t]}_{\text{noise}},
\end{eqnarray}

\noindent where $\varrho_{n}[t] \sim \mathcal{N}(0, \sigma_{n}^{2})$ represents the additive white Gaussian noise characterized by a variance $\sigma_{n}^{2}$ in time slot $t$.

\par Consequently, the transmission rate for the $n$-th user in time slot $t$ is given by
\begin{equation}
    \label{equ_10_1}
    R_{n}[t] = \log_{2}(1+\gamma_{n}[t]),
\end{equation}

\noindent where $\gamma_{n}[t]$ represents the signal-to-interference-plus-noise ratio (SINR) of the $n$-th user in time slot $t$, \textit{i.e.},
\begin{equation}
    \label{equ_11}
    \gamma_{n}[t] = \frac{\left|\widetilde{\bm{h}}_{n}^{\text{bu}}[t]\bm{\omega}_{n}[t]\right|^{2}}{\sum_{i=1, i \neq n}^{N}\left|\widetilde{\bm{h}}_{n}^{\text{bu}}[t]\bm{\omega}_{i}[t]\right|^{2}+\sigma_{n}^{2}}.
\end{equation}

\subsection{Sensing Model}
\par ISAC can use the transmitted communication signals for sensing operations by sharing the spectrum resources~\cite{Xu2024}, thereby achieving efficient resource utilization. Given that the presence of obstacles impacts the sensing performance of the BS, a low-altitude IRS is deployed to reconfigure the sensing link by adjusting the phase of its reflective elements, which enables the direct target sensing services~\cite{Liao2023}. In this case, the received signal at the IRS is given by
\begin{equation}
    \mathbf{x}_{\text{r}}[t] = \bm{\Phi}[t]\bm{H}^{\text{br}}[t]\left(\sum_{n=1}^{N}\bm{\omega}_{n}[t]s_{n}[t]\right).
\end{equation}

\par Accordingly, the covariance matrix is denoted as
\begin{equation}
    \mathbf{X}_{\text{r}}[t] = \bm{\Phi}[t]\bm{H}^{\text{br}}[t]\left(\sum_{n=1}^{N}\bm{\omega}_{n}[t]\bm{\omega}_{n}^{H}[t]\right)(\bm{H}^{\text{br}}[t])^{H}\bm{\Phi}^{H}[t].
\end{equation}

\par In this work, we employ beamforming gain to evaluate the sensing performance of the system, which quantifies the degree of energy concentration in a specific direction. Clearly, a higher beamforming gain indicates better sensing performance of the system. The beampattern gain towards the direction of the target, \textit{i.e.}, the azimuth angle $\vartheta_{\text{ST}}[t]$ and elevation angel $\varphi_{\text{ST}}[t]$, is given by
\begin{equation}
    \label{equ_14}
    \begin{aligned}
    G[t] = \bm{b}^{H}[t]\bm{\Phi}[t]\bm{H}^{\text{br}}[t]\left(\sum_{n=1}^{N}\bm{\omega}_{n}[t]\bm{\omega}_{n}^{\text{H}}[t]\right)(\bm{H}^{\text{br}}[t])^{H}\bm{\Phi}^{H}[t]\bm{b}[t],
    \end{aligned}
\end{equation}

\noindent where $\bm{b}[t]=\bm{\alpha}_{\text{ST}}^{\text{x}}[t] \otimes\bm{\alpha_{\text{ST}}^{\text{y}}}[t]$ denotes the steering vector of the IRS, defined explicitly as~\cite{Hua2024, Liao2023}
\begin{subequations}
\begin{eqnarray}
    \bm{\alpha}_{\text{ST}}^{\text{x}}[t]=[1, e^{j\frac{-2\pi d_{\rm r} \tilde{X}_{\text{ST}}[t]}{\lambda}}, \dots, e^{j\frac{-2\pi (L_{\rm x}-1) d_{\rm r} \tilde{X}_{\text{ST}}[t]}{\lambda}}]^{T}, \\
    \bm{\alpha}_{\text{ST}}^{\text{y}}[t]=[1, e^{j\frac{-2\pi d_{\rm c}\tilde{Y}_{\text{ST}}[t]}{\lambda}}, \dots, e^{j\frac{-2\pi (L_{\rm y}-1)d_{\rm c}\tilde{Y}_{\text{ST}}[t]}{\lambda}}]^{T},
\end{eqnarray}
\end{subequations}

\noindent where $\tilde{X}_{\text{ST}}[t]=\sin\vartheta_{\text{ST}}[t]\cos\varphi_{\text{ST}}[t]$ and $\tilde{Y}_{\text{ST}}[t]=\sin\vartheta_{\text{ST}}[t]\sin\varphi_{\text{ST}}[t]$.

\subsection{UAV Mobility and Energy Model}
\par We assume that the UAV flies at a fixed altitude $z^{\text{r}}$ within the target area for task execution. The horizontal location $\left(x^{\text{r}}[t], y^{\text{r}}[t]\right)$ of the UAV (\textit{i.e.}, low-altitude IRS) in time slot $t$ depends on the flight speed $v_{\text{u}}[t]$ and yaw angle $\theta_{\text{u}}[t] \in [-\pi, \pi]$, which can be denoted as
\begin{eqnarray}
    \label{equ_15}
    \begin{aligned}
    & x^{\text{r}}[t] = x^{\text{r}}[t-1] + v_{\text{u}}[t]t_{\text{d}}\cos(\theta_{\text{u}}[t]), \\
    & y^{\text{r}}[t] = y^{\text{r}}[t-1] + v_{\text{u}}[t]t_\text{d}\sin(\theta_{\text{u}}[t]).
    \end{aligned}
\end{eqnarray}

\par Due to the limited energy resources of the UAV, optimizing propulsion energy consumption can enhance endurance, thereby extending mission execution time and improving the overall operational efficiency of the system. The propulsion energy of the UAV in time slot $t$ can be expressed as~\cite{Dai2022}
\begin{align}
    \label{equ_16}
    E_{\text{u}}[t]&=\left(P_{\text{a}}\left(1+\frac{3v_{\text{u}}^2[t]}{V_{\text{tip}}^{2}}\right) + P_{\text{b}}\left(\sqrt{1+\frac{v_{\text{u}}^{4}[t]}{4v_{\text{a}}^{4}}} -\frac{v_{\text{u}}^{2}}{2v_{\text{a}}^{2}}\right)^\frac{1}{2}\right. \nonumber \\
    &\left. +\frac{1}{2}d_{\text{a}}\rho s A v_{\text{u}}^{3}[t]\right)t_{\text{d}},
\end{align}
\noindent where $P_{\text{a}}$ and $P_{\text{b}}$ are the blade profile power in hovering and induced power, respectively. Moreover, $V_{\text{tip}}$ and $v_{\text{a}}$ represent the tip speed of the rotor blade and mean rotor induced velocity during hovering. In addition, $d_{\text{a}}$, $\rho$, $s$, and $A$ denote the drag ratio, air density, rotor solidity, and disc area, respectively.

\section{Problem Formulation and Analyses}
\label{sec: Problem Analyses}

\par In this section, we aim to formulate the optimization problem that focuses on the communication, sensing, and energy efficiency performance of the considered low-altitude IRS-assisted ISAC system. In the following, the variables and objectives involved in the optimization are first presented, followed by the formulation and analysis of the corresponding problem.

\subsection{Problem Formulation}
\par We begin by specifying the optimization variables as follows: \textit{(i)} $\bm{\Omega}=\{\bm{\omega}_{n}[t]| n\in\mathcal{N}, t\in \mathcal{T}\}$ represents the active beamforming vectors of the BS for all users in all time slots. \textit{(ii)} $\bm{\Psi}=\{\bm{\Phi}[t]| t \in \mathcal{T}\}$ denotes the phase shift matrix of the IRS in all time slots. \textit{(iii)} $\bm{\Lambda}=\{v_{u}[t]|t \in \mathcal{T}\}$ represents the flight velocity of the UAV in all time slots. \textit{(iv)} $\bm{\Theta}=\{\theta_{u}[t]|t \in \mathcal{T}\}$ denotes the yaw angle of the UAV in all time slots.

\par Then, based on the optimization variables above, the optimization objectives are presented as follows:

\par \textit{Objective 1}: Improving communication quality requires maximizing the total transmission rate between the BS and users, which serves as the primary optimization objective. Consequently, our first objective can be expressed as
\begin{equation}
    \label{equ_17}
    f_{1}(\bm{\Omega}, \bm{\Psi},\bm{\Lambda},\bm{\Theta}) = \sum_{t \in \mathcal{T}} R^{\text U}[t],
\end{equation}
\noindent where $R^{\text U}[t]=\sum_{n \in \mathcal{N}}R_{n}[t]$ represents the sum communication rate of all users in time slot $t$.

\par \textit{Objective 2}: Beampattern gain is a key metric for evaluating the sensing performance, which directly decides the quality and reliability of sensing tasks. Thus, the beampattern gain of the target is maximized as the second objective, which can be written as
\begin{equation}
    \label{equ_18}
    f_{2}(\bm{\Omega}, \bm{\Psi},\bm{\Lambda},\bm{\Theta}) = \sum_{t \in \mathcal{T}}G[t].
\end{equation}

\par \textit{Objective 3}: A UAV is an energy-constrained airborne platform, and reducing its energy consumption can extend the operating time of the considered system. Therefore, the third objective focuses on minimizing propulsion energy, expressed as follows:
\begin{equation}
    \label{equ_19}
    f_{3}(\bm{\Lambda}, \bm{\Theta}) = \sum_{t \in \mathcal{T}} E_{\rm u}[t].
\end{equation}

\par Based on the previously defined optimization variables and objectives, we formulate the multi-objective optimization problem as follows:
\begin{subequations}
\label{equ_20}
\begin{align}
    \mathcal{P}: \ &\max_{\bm{\Omega}, \bm{\Psi}, \bm{\Lambda}, \bm{\Theta}}(f_{1}, f_{2}, -f_{3}) \label{Za}\\
    \text{s.t.}: \
    & X_{ \rm min} \leq x^{\rm r}[t] \leq X_{\rm max}, \quad \forall t \in \mathcal{T}, \label{Zb}\\
    &Y_{\rm min} \leq y^{\rm r}[t] \leq Y_{\rm max}, \quad \forall t \in \mathcal{T},  \label{Zc}\\
    &V_{\rm min} \leq v_{\rm u}[t] \leq V_{\rm max}, \quad \forall t \in \mathcal{T},  \label{Zd}\\
    &-\pi \leq \theta_{\rm u}[t] < \pi, \quad \forall t \in \mathcal{T}, \label{Ze}\\
    &\sum_{n=1}^{N}|\bm{\omega}_{n}[t]|^{2} \leq P_{\rm max}, \quad \forall t \in \mathcal{T}, \label{Zf} \\
    &\theta_{l_{x},l_{y}}[t] \in [0, 2\pi), \quad \forall l_{\text{x}} \in \mathcal{L}_{x}, \ l_{\text{y}} \in \mathcal{L}_{y}, \ t \in \mathcal{T}, \label{Zg}
\end{align}
\end{subequations}

\noindent where constraints (\ref{Zb}) and (\ref{Zc}) represent the boundaries of the area within which the UAV performs the mission. Moreover, the flight speed and yaw angle in each time slot are bounded by constraints (\ref{Zd}) and (\ref{Ze}). In addition, constraint (\ref{Zf}) indicates that the transmission power of the BS is less than $P_{\rm max}$. Furthermore, constraint (\ref{Zg}) specifies the phase shift constraint of the IRS.

\subsection{Problem Properties}
\par It can be observed that the optimization problem exhibits the following characteristics:

\par \textit{Non-convexity}: Based on the three optimization objectives presented in Eqs.~(\ref{equ_17})-(\ref{equ_19}), it can be observed that the optimization variables $\bm{\Omega}$, $\bm{\Psi}$, $\bm{\Lambda}$, and $\bm{\Theta}$ are coupled. Moreover, IRS phase shift $\theta_{l_{x},l_{y}}$[t] appears exponentially in the IRS beamforming vector~\cite{Zuo2023}. As a result, the non-convexity of the optimization problem renders it difficult to obtain an optimal solution.

\par \textit{NP-hard}: When only the first optimization objective is considered, optimizing the IRS phase shifts while keeping $\bm{\Omega}$, $\bm{\Lambda}$, and $\bm{\Theta}$ fixed leads to a simplified problem that exhibits the characteristics of a non-convex quadratically constrained quadratic programming (QCQP) problem~\cite{Xie2026}, which is known to be NP-hard~\cite{Wu2021tutorial}. Therefore, the optimization problem also belongs to the NP-hard problem.

\par \textit{Trade-off}: The three optimization objectives exhibit mutual conflicts. For example, focusing on the communication performance will impact the sensing capability of the BS for the target under given channel conditions. Moreover, improving both sensing and communication performance of the system requires rapid deployment of the low-altitude IRS to suitable regions, which increases the UAV propulsion energy consumption. Therefore, there is a trade-off among the three optimization objectives.

\par \textit{Long-term Optimization}: Since the optimization objective is evaluated over $T$ time slots, the solution obtained for an individual time slot does not necessarily correspond to the optimal decision over the entire time horizon. This is because decisions made at any time slot have a continuous and profound impact on the subsequent system state and the feasible decision space. Therefore, the formulated optimization problem inherently involves long-term optimization and demands a trade-off between short-term and long-term objectives.

\section{Proposed Algorithm}
\label{section: Proposed Algorithm}

\par In this section, we first present the motivation behind the adoption of DRL, followed by a transformation of the considered problem into an MDP formulation. Next, we present the core principles of the conventional DDPG algorithm and the improved factors. Finally, we show the proposed GDMDDPG algorithm and analyze it.

\subsection{Motivation of Using DRL}
\par Conventional optimization methods typically handle the problem by decomposing it into several subproblems and applying alternating optimization techniques, which may degrade the precision of the obtained solution~\cite{Zhang2025a}. Moreover, the prior knowledge required by conventional optimization algorithms is difficult to obtain in such a dynamic environment~\cite{Guo2023}. In addition, conventional optimization algorithms typically optimize instantaneous objectives based on the current system state and cannot effectively capture the long-term effects of current decisions, making them unsuitable for solving optimization problems with long-term optimization objectives~\cite{Liang2024}. Furthermore, conventional optimization algorithms typically require iterative optimization in every time slot of the online inference phase in the dynamic environment, making it difficult for them to achieve real-time dynamic decision-making. 

\par In contrast, DRL algorithms interact with the environment based on the trial-and-error learning mechanism without prior knowledge, and possess superior capabilities for balancing short-term and long-term rewards in the dynamic environment. Moreover, trained DRL possesses the capability to make real-time decisions based on the observed environment state during the online inference phase. As such, DRL provides a feasible solution to the aforementioned challenges faced by conventional optimization algorithms. In this case, we propose a GDMDDPG algorithm in the following to solve the formulated optimization problem.

\subsection{MDP Construction}
\par DRL is based on two fundamental components, \textit{i.e.}, the agent and environment. To model the decision-making process of the agent within its environment, the MDP serves as a mathematical foundation. MDP is represented as a five-element structure $\langle \mathcal{S}, \mathcal{A}, \mathcal{R}, \gamma, \mathcal{P} \rangle$, where each element defines a distinct aspect of the agent-environment interaction. Specifically, $\mathcal{S}$ corresponds to the state space, while $\mathcal{A}$ refers to the possible actions the agent can take. The reward function $\mathcal{R}$ quantifies the immediate benefit the agent receives after performing an action in a certain state. Moreover, $\gamma$ denotes the discount factor that determines the relative weighting between future rewards and immediate returns. In addition, $\mathcal{P}$ captures the probabilities governing state transitions after an action is performed. Within the MDP framework, $\mathcal{S}$, $\mathcal{A}$, and $\mathcal{R}$ play pivotal roles, which are explicitly detailed in the context of our formulated optimization problem.

\subsubsection{State Space $\mathcal{S}$}
\par The environment state $\bm{s}[t] \in \mathcal{S}$ observed by the agent in time slot $t$ is defined as $\bm{s}[t]=[t, \bm{q}^{r}[t], \bm{q}^{\rm u}[t], \bm{q}^{\rm g}[t]]$, where $t$ represents the index of current time slot in the pre-defined time slot set $\mathcal{T}$. Moreover, $\bm{q}^{r}[t]$ and $\bm{q}^{\rm g}[t]$ are the locations of the low-altitude IRS and the target in time slot $t$. In addition, $\bm{q}^{\rm u}[t]=[\bm{q}_{1}^{\rm u}[t], \bm{q}_{2}^{\rm u}[t],\ldots,\bm{q}_{N}^{\rm u}[t]]$ is the location set of all users in time slot $t$.

\subsubsection{Action Space $\mathcal{A}$} 
\par According to the observed environment state $\bm{s}[t]$ in time slot $t$, the agent can take the action $\bm{a}[t]\in \mathcal{A}$, which is denoted as $\bm{a}[t]=[\bm{\omega}[t], \bm{\Phi}[t], v_{u}[t], \theta_{u}[t]]$. Specifically, $\bm{\omega}[t]=[\bm{\omega}_{1}[t],\bm{\omega}_{2}[t], \ldots, \bm{\omega}_{N}[t]]$ represents the beamforming vectors of the BS for all users in time slot $t$, and $\bm{\Phi}[t]$ represents the phase shift matrix of the low-altitude IRS in time slot $t$. Moreover, $v_{\text{u}}[t]$ and $\theta_{\text{u}}[t]$ denote the flight velocity and yaw angle of the UAV in time slot $t$, respectively.

\subsubsection{Reward Function $\mathcal{R}$}
\par The reward function is essential in DRL training because it quantifies the quality of the actions taken by the agent, directly influencing the learning direction and strategy selection of the agent. As such, the reward function typically aligns with the optimization objectives, facilitating problem-solving through DRL. Accordingly, the reward function in our considered scenario is expressed as
\begin{equation}
    \label{eq:redesigned reward function}
    r[t] = \eta_{1}R^{\text{U}}[t] + \eta_{2} G'[t] + \eta_{3}E'[t]- PV,
\end{equation}
\noindent where $PV$ is a positive penalty value imposed on the agent when the agent violates the constraints. Moreover, $\eta_{1}$, $\eta_{2}$, and $\eta_{3}$ are constants used to align the magnitudes of the optimization objectives. However, since $G[t]$ and $E[t]$ differ greatly in magnitude and both fluctuate significantly during training, it is difficult and impractical to directly find suitable $\eta_{1}$, $\eta_{2}$, and $\eta_{3}$. In this case, considering that many existing studies express the beampattern gain in decibels (dB), we first use $G'[t]=10\log_{10}G[t]$ to represent the unit conversion of $G[t]$~\cite{Singh2025}. In addition, we utilize $E'[t] = (E_{\text{max}}-E[t])/(E_{\text{max}}-E_{\text{min}})$ to represent the normalization of $E[t]$, where $E_{\text{max}}$ and $E_{\text{min}}$ denote the maximum and minimum values of the UAV energy consumption, respectively. After applying the logarithmic transformation and normalization above, the fluctuation ranges of the beampattern gain and UAV energy consumption are significantly reduced, which makes it easier to find suitable $\eta_{1}$, $\eta_{2}$, and $\eta_{3}$ to keep the three processed optimization objectives (\textit{i.e.}, $R^{\text{U}}[t]$, $G'[t]$, and $E'[t]$) of the same order of magnitude. Furthermore, $G'[t]$ and $E'[t]$ also preserve the effective physical meaning.

\subsection{Principles of DDPG Algorithm}
\par DDPG is a widely used DRL algorithm based on the Actor-Critic architecture and oriented to continuous action space~\cite{Lillicrap2016}, which can be regarded as an extended version of deep Q-network (DQN). The fundamental principles of the DDPG algorithm are presented in detail below.

\subsubsection{Actor-Critic Framework}
\par In the DDPG algorithm, neural networks are leveraged to approximate both the policy and the Q-function. These are represented by the actor network $\bm{\mu}(\bm{s}|\bm{\eta})$ and the critic network $\bm{Q}(\bm{s},\bm{a}|\bm{\zeta})$, where $\bm{\eta}$ and $\bm{\zeta}$ denote the respective parameters of each network. Specifically, the actor network is responsible for generating actions based on the current state, while the critic network estimates the quality of these actions in terms of their expected long-term return. To ensure stability during training, DDPG incorporates target networks. These include the target actor network $\bm {\mu}(\bm{s}|\bm{\eta'})$ and the target critic network $\bm Q(\bm s,\bm a|\bm \zeta')$, where $\bm \eta'$ and $\bm \zeta'$ represent the parameters of the target networks.

\subsubsection{Experience Replay Buffer}
\par The experience replay buffer is set up to improve the sample efficiency of DRL. Specifically, as the agent interacts with the environment, experience tuples $(\bm s[t], \bm a[t], r[t], \bm s'[t])$ are generated and then retained in the buffer, which can be sampled and reused multiple times during subsequent network training, thereby enhancing data utilization. Note that consecutive experiences in the buffer tend to be strongly correlated because they are derived from successive interactions with the environment. However, training the neural network on such highly correlated data may result in unstable convergence. In this case, randomly sampling experiences from the buffer and performing network updates can increase diversity and reduce correlations of the training data, which improves the training stability.

\subsubsection{Network Update}
\par The training objective of the critic network in DDPG is to minimize the temporal difference (TD) error, which measures the discrepancy between the target Q-value and the current Q-value, denoted as
\begin{equation}
    \label{equ_25}
    L(\bm \zeta) = \mathbb{E}\left[\left(\bm Q\left(\bm s[t],\bm a[t]|\bm \zeta\right)-y[t]\right)^{2}\right], 
\end{equation}
\noindent where $y[t]$ is the target value.

\par We can approximate the loss function in Eq.~(\ref{equ_25}) by randomly sampling $B$ experiences from the replay buffer for updates, which is given by
\begin{equation}
    \label{equ_26}
    L(\bm \zeta) = \frac{1}{B}\sum_{b=1}^{B}\left(\bm Q\left(\bm s_{b},\bm a_{b}|\bm \zeta\right)-y_{b}\right)^{2}, 
\end{equation}
\noindent where $y_{b}=r_{b}+\gamma \bm Q(\bm s_{b}', \bm \mu(\bm s_{b}'|\bm \eta')|\bm \zeta')$ is the target Q-value

\par Note that the actor network adjusts its actions based on the evaluation of the critic network. As such, the actor network can be updated as follows:
\begin{equation}
    \label{equ_28}
    L(\bm \eta) = -\frac{1}{B}\sum_{b=1}^{B}\left(\bm Q(\bm s_{b}, \bm \mu(s_{b}|\bm \eta)|\bm \zeta)\right).
\end{equation}

\par To mitigate training instability caused by significant variations in the target networks, a soft-update strategy is adopted for updating the target networks as follows:
\begin{eqnarray}
    \label{equ_29}
    \bm \eta' \leftarrow \varepsilon \bm \eta + (1 - \varepsilon)\bm \eta', \quad \bm \zeta' \leftarrow \varepsilon \bm \zeta + (1-\varepsilon) \bm \zeta',
\end{eqnarray}
\noindent where $\varepsilon$ is a small constant that controls the soft-update rate.

\subsection{GDMDDPG}
\label{Sec: Proposed Improved Solution}
\par In this part, we first analyze the limitations of the conventional DDPG algorithm, and then introduce the principles of the corresponding improved factors.

\subsubsection{The Limitations of Conventional DDPG}

\par First, in the conventional DDPG algorithm, the actor network typically employs MLP as the core architecture due to the simplicity and ease of implementation. However, in more complex environments, the fixed MLP architecture may fail to fully capture and analyze the intricate information present in the environment states, which in turn limits the decision-making capabilities of the agent~\cite{Du2024a}. Notably, the diffusion model provides a viable solution to address the above limitations due to its strong ability in modeling complex data distributions and generating high-quality samples~\cite{10839238}. By replacing the traditional MLP with the diffusion model, the actor network can more effectively analyze the environment states, thereby generating superior actions and further enhancing the overall system performance.

\par Second, DDPG is based on a deterministic policy, where the actor network directly outputs a specific action value rather than a probability distribution over actions. Such a deterministic policy constrains the agent in achieving sufficient exploration over the action space~\cite{Colas2018}. To address this issue, noise can be added to the actions generated by the actor network, introducing randomness to promote the agent exploration of the action space. Moreover, this noise perturbation mechanism helps alleviate the tendency of the DDPG to output action boundary values during the early training phase, thereby enhancing the convergence speed of the training process.

\par Finally, the experience replay buffer in conventional DDPG employs a random sampling strategy, which might fail to prioritize experiences that hold significant learning value, resulting in slower convergence~\cite{Goek2024}. To address this issue, the RPER sampling mechanism, which integrates ERE and PER techniques, can be incorporated into the network update process of the DDPG. Specifically, the ERE technique ensures that more recent experiences are sampled more actively. Based on this, the PER technique assigns a priority to each experience sample, allowing those that contribute more to learning to be sampled with higher probability. As such, combining PER and ERE can effectively mitigate the limitations of the random sampling strategy.

\par The diffusion model-based actor network, noise perturbation mechanism, and RPER sampling mechanism are described in detail below.

\subsubsection{Diffusion Model-based Actor Network}
\par The forward process and reverse process are two key processes in the diffusion~\cite{Yang2024Diffusion}. In the former, the model gradually perturbs the original data by adding noise step by step, transforming the data into a nearly pure noise state. In the latter, the model learns the inverse of the noise-adding process, attempting to progressively denoise the noisy data and ultimately recover the original data. In particular, this progressive denoising process enables the model to capture complex nonlinear relationships and dependencies in the data, thus enabling deep modeling of the data distribution. As such, diffusion model-based DDPG is better able to generate progressively more refined and optimized decisions based on the current environment state during exploration and exploitation, and enhance its adaptability in complex environments. Next, the detailed mathematical representation of the diffusion model is presented.

\par \textbf{\textit{Forward Process}}: We start from the original data $x_0$ and define a forward noise-injection process composed of $G$ discrete steps, indexed by $\mathcal{G} = \{1, \ldots, g, \ldots, G\}$. At each step $g \in \mathcal{G}$, the perturbed data is represented as $\bm x_g$, which shares the same shape as $\bm x_0$. Throughout this process, Gaussian noise is gradually introduced, producing a sequence of increasingly noisy samples. This leads to a Markov chain $\{\bm x_0, \bm x_1, \ldots, \bm x_G\}$, where each transition from $\bm x_{g-1}$ to $\bm x_g$ is governed by a conditional distribution $\bm q(\bm x_g \mid \bm x_{g-1})$, defined as
\begin{equation}
    \label{equ_31}
    \bm q(\bm x_{g} | \bm x_{g-1}) = \mathcal{N}(\bm x_{g}; \sqrt{1 - \beta_{g}} \, \bm x_{g-1}, \beta_{g} \mathbf{I}),
\end{equation}
\noindent where $\mathbf{I}$ is the identity matrix, and $\beta_{g}$ denotes the noise variance at the $g$-th step, which can be calculated according to the Variational Posterior (VP) scheduler as follows:
\begin{equation}
    \label{equ_32}
    \beta_{g}=1-e^{-\frac{{c_1}}{G}-\frac{2g-1}{2G^{2}}({c_2}-{c_1})},
\end{equation}
\noindent where ${c_1}$ and ${c_2}$ are constant parameters.

\par From Eq.~(\ref{equ_31}), it can be observed that $\bm x_{g} $ depends only on $\bm x_{g-1}$, which allows the forward process to be modeled as a Markov chain. In this case, the transition distribution $\bm q(\bm x_{G}|\bm x_{0})$ from the original data $\bm x_{0}$ to $\bm x_{G}$ can be expressed as
\begin{equation}
    \label{equ_33}
    \bm q(\bm x_{G}|\bm x_{0}) = \prod_{g=1}^{G}\bm q(\bm x_{g}|\bm x_{g-1}).
\end{equation}

\par However, when the value of $g$ is large, sampling $\bm x_{g}$ directly according to Eq.~(\ref{equ_33}) becomes computationally intensive and time-consuming. In this case, based on the mathematical relationship between $\bm x_{0}$ and $\bm x_{g}$, $\bm x_{g}$ can be expressed as:
\begin{equation}
    \label{equ_34}
    \bm x_{g} = \sqrt{{\hat{\alpha}_{g}}}\bm x_{0} + (\sqrt{1-{\hat{\alpha}_{g}}})\bm \epsilon,
\end{equation}
\noindent where $\alpha_{g}=1-\beta_{g}$, and $\hat{\alpha}_{g}=\prod_{i=1}^{g}\alpha_{i}$ is the product of all $\alpha_i$ ($i\leq g$). Moreover, $\bm \epsilon \sim \mathcal{N}(0, \mathbf{I})$ represents the standard normal noise.

\par Note that the forward process is based on the original data $\bm x_{0}$, which can be regarded as the optimal solution to the optimization problem. However, in our considered dynamic ISAC scenario, it is impractical to obtain the optimal solution in advance~\cite{Du2024a}. Therefore, only the subsequent reverse process is employed to improve the performance of the DDPG actor network.

\begin{figure*}[t]
    \centering
    \includegraphics[width=\linewidth]{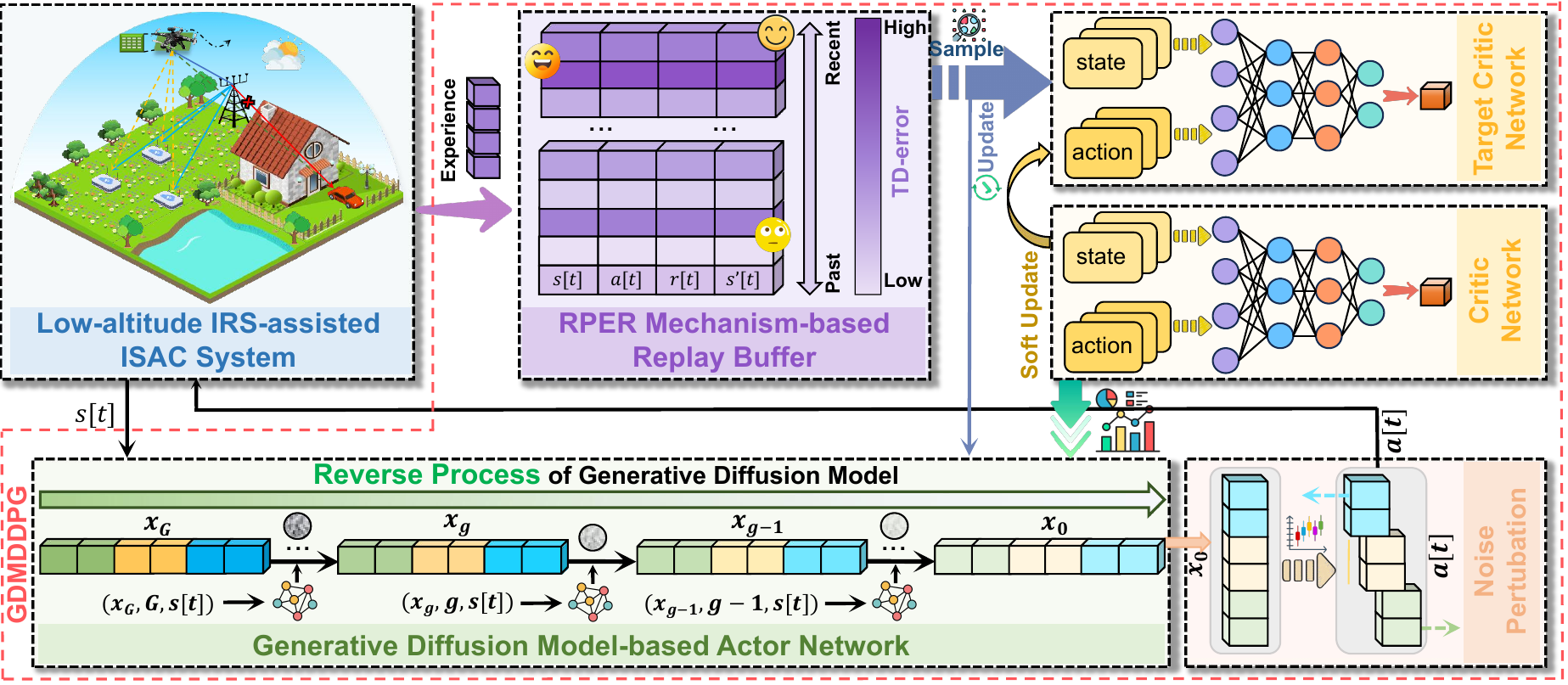}
    \caption{The architecture of the proposed GDMDDPG algorithm. Specifically, the GDM is integrated into the actor network to make decisions, which is perturbed by the noise perturbation mechanism. Moreover, the RPER mechanism is used to improve the learning efficiency.}
    \label{fig: structure of the proposed algorithm}
\end{figure*}

\par \textit{\textbf{Reverse Process}}: The reverse process is formulated as a sequential denoising procedure that reconstructs the original sample $\bm{x}_0$ from the noisy input $\bm{x}_G$. Ideally, if the true reverse conditional distribution $\bm p(\bm{x}_{g-1}|\bm{x}_g)$ were available for all $g \in \{G, G{-}1, \dots, 1\}$, one could recover $\bm x_0$ by sampling through the full reverse trajectory. However, exact evaluation of $\bm p(\bm x_{g-1} \mid \bm x_g)$ is generally intractable due to the complexity of the data distribution. To avoid this issue, a parameterized distribution $\bm{p}_{\bm{\eta}}(\bm{x}_{g-1}|\bm {x}_g)$ is introduced to approximate the reverse process, which can be defined as
\begin{equation}
    \label{equ_35}
    \bm p_{\bm \eta}(\bm x_{g-1}|\bm x_{g}) = \mathcal{N}(\bm x_{g-1}; \bm \kappa_{\bm \eta}(\bm x_{g}, g, \bm d), \tilde{\beta}_{g}\mathbf{I}),
\end{equation}
\noindent where $\bm d$ represents the condition information. Moreover $\tilde{\beta}_{g}\mathbf{I}$ is the variance of the distribution, and $\tilde{\beta}_{g}$ can be represented as
\begin{equation}
    \label{equ_36}
    \tilde{\beta}_{g} = \frac{1-\hat{\alpha}_{-1}}{1-\hat{\alpha}_{g}}\beta_{g}.
\end{equation}
\par The mean $\bm \kappa_{\bm \eta}(\bm x_{g}, g, \bm d)$ is a deep model with parameter $\eta$. Note that $\bm \kappa_{\bm \eta}(\bm x_{g}, g, \bm d)$ can be expressed as
\begin{equation}
    \label{equ_37}
    \bm \kappa_{\bm \eta}(\bm x_{g}, g, \bm d) = \frac{\sqrt{\alpha_{g}}(1-\hat{\alpha}_{g-1})}{1-\hat{\alpha_{g}}}x_{t} + \frac{\sqrt{\hat{\alpha}_{g-1}}\beta_{g}}{1-\hat{\alpha}_{g}}\bm x_{0}.
\end{equation}

\par Moreover, it can be observed that $\bm x_{0}$ can be obtained according to Eq.~(\ref{equ_34}), which can be represented as
\begin{equation}
    \label{equ_38}
    \bm x_{0} = \frac{x_{g}}{\sqrt{\hat{\alpha}_{g}}}-\frac{{\sqrt{1-\hat{\alpha}_{g}}}\text{tanh}(\bm \epsilon_{\bm \eta}(\bm x_{g}, g, \bm d))}{\sqrt{\hat{\alpha}_{g}}},
\end{equation}
\noindent where $\bm \epsilon_{\bm \eta}(\bm x_{g}, g, \bm d)$ is a deep neural network with parameter $\bm \eta$, and its role is to generate the denoising noise at the $g$-th step based on the condition information $\bm d$. Moreover, to prevent the denoising noise generated by $\bm \epsilon_{\bm \eta}$ from becoming too large and affecting the actions, the tanh function is used to scale it. However, Eq.~(\ref{equ_38}) cannot be directly used to generate $\bm x_0$, because $\bm \epsilon_{\bm \eta}$ is not the same as $\bm \epsilon$ in Eq.~(\ref{equ_34}). Instead, Eq.~(\ref{equ_38}) can be applied to Eq.~(\ref{equ_37}) to estimate the mean of the distribution, which can be denoted as
\begin{equation}
    \label{equ_39}
    \bm \kappa_{\bm \eta}(\bm x_{g}, g, \bm d) = \frac{1}{\sqrt{\alpha_{g}}}\left(\bm x_{g}-\frac{\beta_{g}}{\sqrt{1-\bar{\alpha}_{g}}}\text{tanh}\left(\bm \epsilon_{\bm \eta}\left(\bm x_{g}, g, \bm d\right)\right)\right).
\end{equation}

\par As such, according to Eqs.~(\ref{equ_39}) and (\ref{equ_36}), the transition distribution $\bm p_{\bm \eta}$ is obtained. Then, the original data $\bm x_{g-1}$ can be sampled from the distribution $\bm p(\bm x_{g})\bm p_{\bm \eta}(\bm x_{g-1}|\bm x_{g})$. Similar to the forward process, the original data $\bm x_{0}$ can be obtained from the distribution as follows:
\begin{equation}
    \label{equ_40}
    \bm p_{\bm \eta}(\bm x_{0}) = \bm p(\bm x_{G})\prod_{g=1}^{G}(\bm p_{\bm \eta}(\bm x_{g-1}|\bm x_{g})),
\end{equation}
\noindent where $\bm p(\bm x_{G})$ is a Gaussian distribution.

\par Therefore, the reverse process of the diffusion model can be integrated into the actor network of the DDPG algorithm, sampling from the distribution $\bm p_{\bm \theta}(\bm x_{0})$ to generate high-quality action. However, during the network training process, the inability to backpropagate gradients through the random variables presents a challenging issue. In this case, the reparameterization method can be used to decouple the random variable from the distribution~\cite{Du2024a}, which is represented as
\begin{equation}
    \label{equ_41}
    \bm x_{g-1} = \bm \kappa_{\bm \eta}(\bm x_{g}, g, \bm s[t]) + (\tilde{\beta}_{g})^{2} \odot \bm \epsilon,
\end{equation}
\noindent where $\bm s[t]$ represents the environment state in time slot $t$, and $\bm \epsilon \sim \mathcal{N}(0, \mathbf{I})$. As such, by iteratively applying Eq.~(\ref{equ_41}), all $\bm x_{g} \ (1 \leq g \leq G)$, and in particular $\bm x_{0}$, \textit{i.e.}, the action $\widetilde{\bm a}[t]$, based on the observed environment state $\bm s[t]$.

\subsubsection{Noise Perturbation Mechanism}
\par Drawing from the principles of the twin delayed deep deterministic policy gradient (TD3) algorithm, we integrate a noise perturbation mechanism to improve the exploration ability of the agent. In this mechanism, the decision $\widetilde{\bm a}[t]$ made by the diffusion model-based actor network at time step $t$ is modified by adding noise to generate the actual action $\bm a[t]$ as follows:
\begin{equation}
    \label{equ_42}
    \bm a[t] = \widetilde{\bm a}[t] + \iota_{a}, \quad \iota_{a} \sim \text{clip}\left(\mathcal{N}\left(0,\hat{\sigma}\right), -b, b\right),
\end{equation}
\noindent where $\iota_{a}$ represents the noise added to the action, which is sampled from a normal distribution with a mean of 0 and a standard deviation of $\hat{\sigma}$. The clip operation restricts the noise to a range of $[-b, b]$, which prevents instability in the policy due to over-exploration.

\begin{algorithm}[t]
    \SetAlgoLined 
	\caption{GDMDDPG}
        \label{Algorithm1}
    \LinesNumbered
    Initialize the weights of the actor and critic networks $\bm \eta$ and $\bm \zeta$. Moreover, Initialize the target networks, \textit{i.e.}, $\bm \eta' \leftarrow \bm \eta$, $\bm \zeta' \leftarrow \bm \zeta$. \\
    Initialize experience replay buffer $\mathcal{B}$, batch size $B$.\\
    Initialize the maximum number of episodes $N_{e}$ and time slot length $T$.\\
    \For{episode = \rm{1 to} $N_{e}$}{
        \For{t = \rm{1 to} $T$}{
            Observe the state $s[t]$ and initialize a random distribution $\bm x_{G} \sim \mathcal{N}(0,1)$. \\
            \For{g = \rm{$G$ to} 1}{
                Deploy a denoising network $\bm \epsilon_{\bm \eta}(\bm x_{g},g,\bm s[t])$. \\
                Calculate the mean $\bm \kappa_{\bm \eta}$ and variance $\tilde{\beta}_{g}$ based on Eqs.~(\ref{equ_39}) and~(\ref{equ_36}).\\
                Calculate $\bm x_{g}$ according to Eq.~(\ref{equ_41}). \\
            }
            Obtain the action $\tilde{\bm a}[t]$ based on the reverse process above, and obtain the action $\bm a[t]$ according to Eq.~(\ref{equ_42}). \\
            Take the action $\bm a[t]$, obtain reward $r[t]$ and next state $\bm s'[t]$.\\
            Store the transition $\{\bm s[t],\bm a[t],r[t],\bm s'[t]\}$ in to the replay buffer $\mathcal{B}$.\\
            Sample $B$ transitions from $\mathcal{B}$ according to the probability in Eq.~(\ref{equ_45}), and then update $\bm \eta$ and $\bm \zeta$ according to Eqs.~(\ref{equ_28}) and~(\ref{equ_47}). \\
            Update the target networks according to Eq.~(\ref{equ_29}). \\
        }
    }
    Return the actor network.
\end{algorithm}

\subsubsection{RPER Sampling Mechanism}
\par To enhance the convergence speed and accuracy of the DDPG algorithm in the training process, we adopt a RPER sampling mechanism to sample the experiences from the replay buffer for network update~\cite{Wang2019}. Note that the RPER sampling mechanism consists of two phases, \textit{i.e.}, defining sampling range and priority-based sampling.

\par \textbf{\textit{Defining Sampling Range}}: Considering that more recent experiences in the experience replay buffer are typically more informative than earlier ones, the sampling range can be gradually reduced during the update process, and increase the sampling frequency of these recent experiences. Specifically, assume that during the current network update phase, a total of $U$ mini-batch updates are required. At the $u$-th update, $1 \leq u \leq U$, the most recent $F_{u}$ experiences are selected from the experience buffer for subsequent sampling, which can be represented as
\begin{equation}
    \label{equ_43}
    F_{u} = \max\{B_{\rm max} \cdot \rho^{(u\cdot\frac{1000}{U})}, F_{\rm min} \},
\end{equation}
\noindent where $B_{\rm max}$ is the capacity of the replay buffer, and $\rho \in (0, 1]$ is a parameter that represents the importance given to recent experiences. Moreover, $F_{\rm min}$ refers to the minimum value within the range of $F_{u}$. It can be observed that this technique allows for frequent sampling of recent experiences while also taking past experiences into account.

\par \textbf{\textit{Priority-based Sampling}}: Given that different experiences contribute variably to the training process, priority-based sampling can enhance training efficiency more effectively than simple random uniform sampling. In the RPER sampling mechanism, the priority of each experience sample is quantified by TD error, and experiences are sampled from the sampling range defined by Eq.~(\ref{equ_43}). In this case, the priority of the $i$-th experience can be expressed as
\begin{equation}
    \label{equ_44}
    p_{i} = |\delta_{i}| + \epsilon_{p}, \quad i \in D_{F_{u}},
\end{equation}
\noindent where $\delta_{i}$ represents the TD error, and $\epsilon_{p}$ is a small positive constant used to prevent the priority of certain experiences from being zero. Moreover, $D_{F_{u}}$ represents the experience set obtained according to Eq.~(\ref{equ_43}).

\par As such, the probability of sampling the $i$-th experience can be denoted as
\begin{equation}
    \label{equ_45}
    P_{i} = \frac{p_{i}^{\beta_{1}}}{\sum_{j \in D_{F_{u}}}p_{j}^{\beta_{1}}}, \quad i \in D_{F_{u}},
\end{equation}
\noindent where $\beta_{1}$ is a hyperparameter that controls the influence of priority on the sampling probability.

\par However, frequent sampling of experiences with high TD errors may lead to instability in the training process. To address this issue, a sampling-importance weight is introduced, which can be expressed as
\begin{equation}
    \label{equ_46}
    c_{i} = (\frac{1}{B_{\rm max}}\cdot\frac{1}{P_{i}})^{\beta_{2}},
\end{equation}
\noindent where $\beta_{2}$ is a hyperparameter that controls the correction extend.

\par Therefore, the loss function of the critic network of the DDPG algorithm in Eq.~(\ref{equ_26}) can be rewritten as follows:
\begin{equation}
    \label{equ_47}
    L(\bm \zeta) = \frac{1}{B}\sum_{b=1}^{B}c_{b}\left(\bm Q\left(\bm s_{b},\bm a_{b}|\bm \zeta\right)-y_{b}\right)^{2}.
\end{equation}

\subsection{Analyses of GDMDDPG algorithm}
\par The structure and comprehensive steps of the GDMDDPG algorithm are presented in Fig.~\ref{fig: structure of the proposed algorithm} and Algorithm~\ref{Algorithm1}, respectively. In the following, we provide the computation complexity analysis and convergence analysis for GDMDDPG.

\subsubsection{Computation Complexity Analysis}

\par The computation complexity of the proposed GDMDDPG algorithm is $\mathcal{O}(2|\bm \eta|+2|\bm \zeta|+N_{e}GT|\eta|+N_{e}TV+N_{e}T(2|\bm \eta|+2|\bm \zeta|+N_{e}T\log F_{u})$. The details are as follows:

\par \textit{Network Initialization}: GDMDDPG algorithm contains an actor network, a critic network, a target actor network, and a target critic network. Assume that the number of parameters for the actor network and the critic network are $|\bm \eta|$ and $|\bm \zeta|$. Moreover, the target network and the main network have the same number of parameters. Therefore, the computation complexity of the network initialization process is $\mathcal{O}(2|\bm \eta|+2|\bm \zeta|)$.

\par \textit{Action Sampling}: The computation complexity of the action sampling process consists of two parts, \textit{i.e.}, the diffusion model-based action generation and noise perturbation mechanism. Assume that the number of training episodes, diffusion steps, and time slots are $N_{e}$, $G$, and $T$, respectively. Then, the computation complexity of the diffusion model-based generated action process is $\mathcal{O}(N_{e}GT|\bm \eta|)$. Moreover, the noise perturbation mechanism applies a Gaussian perturbation directly to the sampled action phase through a simple random number generation followed by vector addition. The computational cost of this mechanism is extremely small, which is negligible in terms of overall complexity. As such, the computation complexity of the action sampling process is $\mathcal{O}(N_{e}GT|\bm \eta|)$.

\par \textit{Experience Collection}: Assume that the complexity of the agent interacting with the environment is $V$, the complexity of the experience collection process is $\mathcal{O}(N_{e}TV)$.

\par \textit{RPER Mechanism-based Experience Sampling and Update}: The computational complexity of the RPER mechanism stems from the ERE mechanism and the PER mechanism. In the ERE mechanism, each update only requires computing the number of most recent experiences $F_{u}$, and the associated computational cost is negligible. In the PER mechanism, a sum-tree data structure is typically used to store experience priorities, which leads to a computation complexity of $\mathcal{O}(\log F_{u})$ for both sampling and priority updates. As such, the computation complexity of the RPER mechanism is $\mathcal{O}(N_{e}T\log F_{u})$.

\par \textit{Network Update}: The network update process consists of the actor and critic network updates and the soft updates of the target networks. Therefore, the computation complexity of the network update process is $\mathcal{O}(N_{e}T(2|\bm \eta|+2|\bm \zeta|))$.

\subsubsection{Convergence Analysis}
\par Providing the rigorous theoretical convergence behavior analysis of the proposed algorithm presents fundamental challenges~\cite{Oubbati2022}. Specifically, the underlying low-altitude IRS-assisted ISAC environment exhibits complex stochastic dynamics with time-varying channel conditions and UAV mobility that defy closed-form analytical characterization. Moreover, the iterative stochastic denoising and sampling mechanism inherent to the generative diffusion model, together with the noise perturbation mechanism applied at the actions, introduces multiple sources of non-stationarity, which makes it extremely difficult to conduct a strict theoretical convergence analysis. Consequently, the proposed algorithm operates in an environment where empirical evaluation provides more meaningful insights than theoretical bounds that would require overly restrictive assumptions. As such, we provide comprehensive convergence trend and stability analysis to demonstrate algorithm convergence capability in the environment, which is shown in Section~\ref{subsubsec:Convergence Performance Evaluation}.

\section{Simulation Results}
\label{sec: Simulations}

\par In this section, we construct simulations to validate the performance of the proposed GDMDDPG algorithm. 

\subsection{Simulation Setup}
\subsubsection{Environment Details}
\par We define the target area as a $300 \ \text{m} \times 300 \ \text{m}$ square region, which includes one BS with multiple antennas that simultaneously serves three communication users and one sensing target, as well as one low-altitude IRS with 16 reflective elements. The total runtime of the system is $100 \ \text{s}$, and the length of every time slot is $1 \ \text{s}$. Moreover, the initial position of the UAV (\textit{i.e.}, low-altitude IRS) is $[0 \ \text{m}, 300 \ \text{m}, 40 \ \text{m}]$, and the UAV maintains a constant altitude during the flight. Other parameters related to the considered scenario are shown in Table~\ref{tab:parameter}.

\begin{table}[t]
\caption{Simulation Parameters}
\label{tab:parameter}
\centering
\setlength{\tabcolsep}{4.5mm}{
\begin{tabular}{@{}@{\extracolsep{\fill}}p{3mm}@{}|llll@{}}
\toprule
 & Symbol & Value & Symbol & Value \\
\midrule
\multirow{10}{*}{\rotatebox[origin=c]{90}{System Model}} & $X_{\rm\min}$ & 300 m & $Y_{\rm\max}$ & 300 m \\
& $V_{\rm max}$ & 30 m/s & $L_0$ & 0.001 \\
& $M$ & 4 & $\eta^{\rm bu}$ & 1 \\
& $\iota^{\rm br}$ & 2.2 & $\iota^{\rm ru}$ & 2.2 \\
& $\iota^{\rm bu}$ & 4.6 & $\sigma^2$ & -90 dBm \\
& $P_{\rm s}$ & 79.85 & $P_{\rm m}$ & 88.63 \\
& $U_{\rm r}$ & 120 m/s & $V_{\rm h}$ & 4.03 \\
& $d_0$ & 0.6 & $\rho_{\rm a}$ & 1.225 kg/m$^3$ \\
& $z$ & 0.05 & $G$ & 0.503 m$^2$ \\ \midrule

\multirow{3}{*}{\rotatebox[origin=c]{90}{DRL}}
& $\xi_{1}$ & 1000 & $p_{\rm o}$ & 50 \\
& $N_{\rm e}$ & 4500 & $\gamma$ & 0.99 \\
& $B_{\max}$ & 100000 & $B$ & 128 \\
& $F_{\min}$ & 3000 & $\varepsilon$ & 0.005 \\ 
\bottomrule
\end{tabular}}
\end{table}

\begin{figure}
    \centering
    \includegraphics[width=\linewidth]{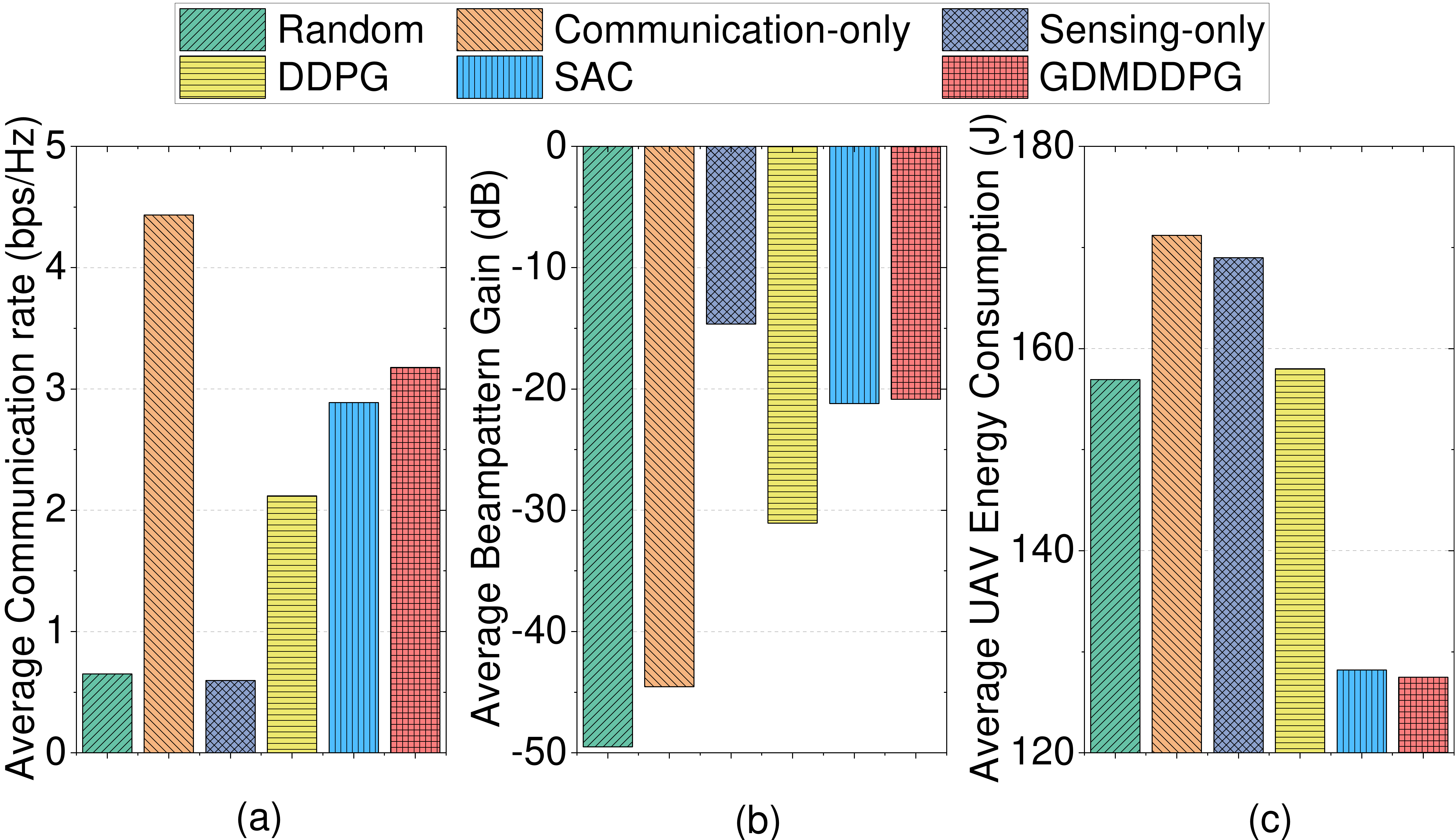}
    \caption{Optimization objective values obtained by different methods. (a) Average communication rate. (b) Average beampattern gain. (c) Average UAV energy consumption.}
    \label{fig: optimization objective values}
\end{figure}

\subsubsection{Network Structure}
\par The actor network takes as input the previous step action, the current reverse step index, and the observed environment state. First, the reverse step index is first transformed into a sinusoidal positional embedding to model temporal dynamics and is then processed through two fully connected layers with 16 and 8 neurons, using ReLU activation in the first layer. Subsequently, the previous action, the reverse step embedding, and the observed state are concatenated and passed through three fully connected layers, where ReLU activation is applied to all layers except the final one. For value estimation, the critic network uses two hidden layers with 256 neurons each and ReLU activation to enhance nonlinear representation. Moreover, the target actor and critic networks adopt the same architecture as their respective main networks to maintain consistency.

\begin{figure*}[t]
    \centering
    \includegraphics[width=0.85\linewidth]{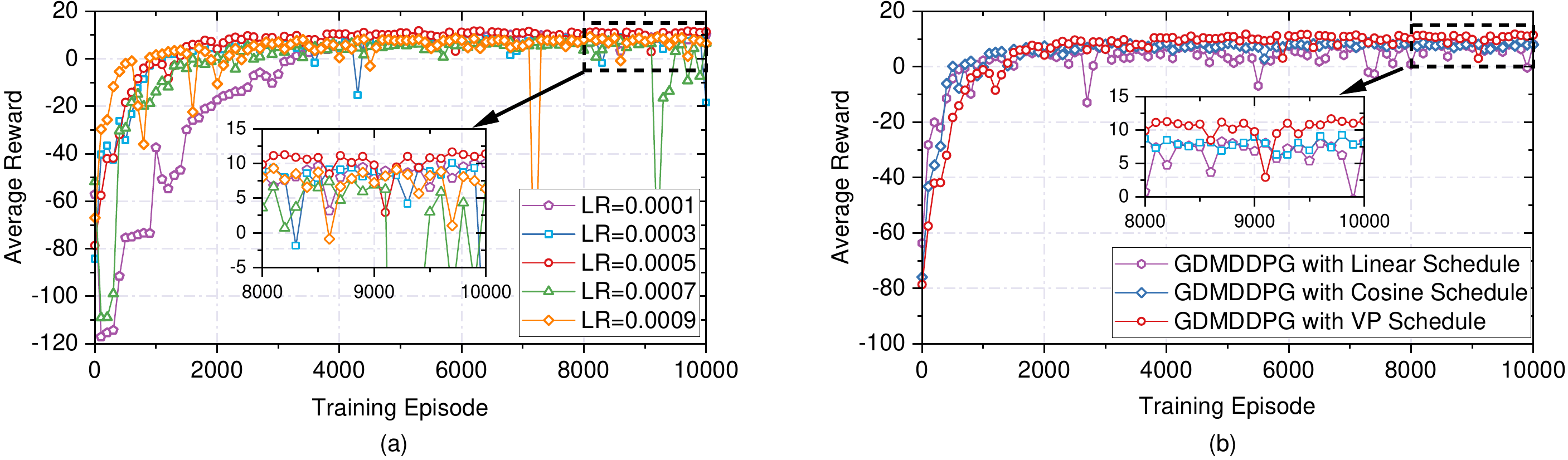}
    \caption{Convergence of the proposed GDMDDPG algorithm with different parameters. (a) Learning rates. (b) Noise schedule functions.}
    \label{fig: LR and noise schedule functions}
\end{figure*}

\subsection{Comparison Methods}

\par The proposed algorithm is evaluated through comparisons with the following methods.:
\begin{itemize}
    \item \textit{Random Method}: In this method, the active beamforming matrix of the BS, IRS phase shifts, and UAV trajectory are set randomly.
    \item \textit{Communication-only Method}: In this method, the GDMDDPG algorithm focuses only on the communication performance of the system, \textit{i.e.}, it only maximizes the communication rate between the BS and the users.
    \item \textit{Sensing-only Method}: In this method, the GDMDDPG algorithm focuses only on the sensing performance \textit{i.e.}, it only maximizes the beampattern gain of the target.
\end{itemize}

\par Moreover, we compare our proposed GDMDDPG algorithm with the following state-of-the-art DRL algorithms:
\begin{itemize}
    \item \textit{DDPG}: The DDPG algorithm is used to solve the optimization problem to verify the effectiveness of the proposed improvement factors. Note that DDPG is applied with the noise perturbation mechanism to encourage exploration~\cite{Lillicrap2016}.
    \item SAC: The SAC algorithm improves policy exploration by maximizing entropy while optimizing the expected cumulative discounted reward, which enables it to achieve strong and stable performance across various tasks~\cite{Haarnoja2018}.
\end{itemize}

\subsection{Optimization Results}
\par We analyze the performance of GDMDDPG under different setup conditions versus the comparison methods.

\subsubsection{Algorithm Performance Evaluation}
\par Fig.~\ref{fig: optimization objective values} presents the optimized objective values obtained by different algorithms. We can find that the proposed GDMDDPG algorithm achieves superior performance compared with the conventional DDPG algorithm, SAC algorithm, and random method in both communication rate and beampattern gain due to its improved environmental understanding and exploration capabilities. Moreover, GDMDDPG demonstrates the best performance in energy consumption optimization, which is crucial for the energy-constrained UAV platform. In particular, the proposed GDMDDPG achieves a 50\% improvement in communication rate and a 32.9\% increase in beampattern gain, while simultaneously reducing UAV energy consumption by 19.3\% compared to the conventional DDPG algorithm. As such, the superior communication and sensing performance combined with the lowest energy consumption establishes GDMDDPG as an applicable and effective algorithm for the low-altitude IRS-assisted ISAC system. In addition, the communication-only and sensing-only methods achieve the best values in communication rate and beampattern gain metrics compared to GDMDDPG, respectively, which confirms that a significant trade-off relationship exists in the optimization objectives in the formulated optimization problem.

\subsubsection{Convergence Performance Evaluation}
\label{subsubsec:Convergence Performance Evaluation}
\par We analyze the impact of different key hyperparameters on the convergence of the proposed GDMDDPG algorithm and further compare its convergence with that of different DRL methods.

\par \textit{(i) Hyperparameter Optimization Results}: Fig.~\ref{fig: LR and noise schedule functions}(a) shows the convergence performance of the GDMDDPG algorithm under different learning rates. The algorithm demonstrates enhanced performance in both accuracy and stability when employing a learning rate of $5 \times 10^{-4}$. This is because, with a smaller learning rate, the learning process of the agent becomes too slow, often leading to premature convergence to a local optimum. On the contrary, a larger learning rate makes the parameter update step of the neural network larger, resulting in the algorithm oscillating or even not converging during the training process.

\par Fig.~\ref{fig: LR and noise schedule functions}(b) shows the convergence performance of the GDMDDPG algorithm under different noise scheduler functions. The effectiveness of the diffusion model is highly dependent on the noise schedule function, which determines how the noise level changes with the diffusion steps. Specifically, this function influences the ability of the model to learn and generate high-quality samples by controlling the noise variation in the diffusion process. To systematically assess the effect of different noise scheduler functions on the proposed GDMDDPG algorithm, we conduct comparative simulations using three representative scheduling functions, \textit{i.e.}, VP, linear, and cosine. As can be seen, the GDMDDPG algorithm achieves superior performance when VP is employed as the noise scheduling function, indicating its greater effectiveness for addressing the formulated optimization problem.

\par
\begin{figure*}[t]
    \centering
    \includegraphics[width=\textwidth]{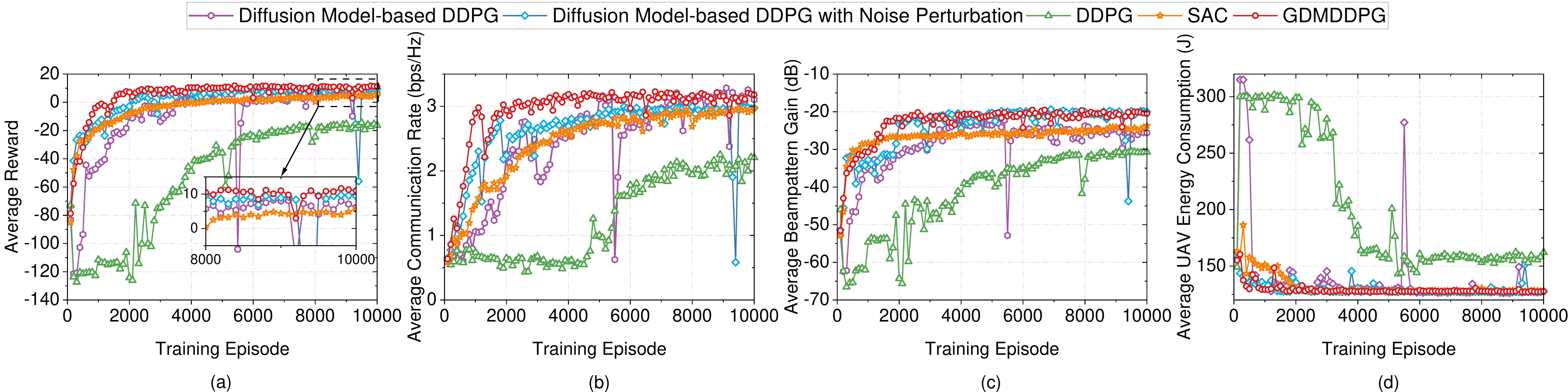}
    \caption{Convergence curves of different methods. (a) Average reward. (b) Average communication rate. (c) Average beampattern gain. (d) Average UAV energy consumption.}
    \label{fig:comparison results}
\end{figure*}

 \textit{(ii) Comparison with Other Methods}: Fig.~\ref{fig:comparison results} shows the convergence curves of all the methods in terms of the reward and three optimization objectives. As we can see, the conventional DDPG algorithm shows the weakest performance in both convergence speed and accuracy due to its limited exploration capability, limited analytical capability, and reduced sample efficiency. Moreover, the SAC algorithm outperforms the DDPG algorithm due to its enhanced exploration performance. In addition, compared to the conventional DDPG algorithm, the diffusion model-based DDPG algorithm achieves a significant improvement, as the diffusion model improves the environment state modeling capability and increases the diversity and quality of decisions. Furthermore, we can find that the noise perturbation mechanism is capable of further improving the decision accuracy of the diffusion model-based DDPG algorithm by strengthening exploration while reducing frequent boundary-value explorations in the early training stage.

 \begin{figure}[t]
    \centering
    \includegraphics[width=\linewidth]{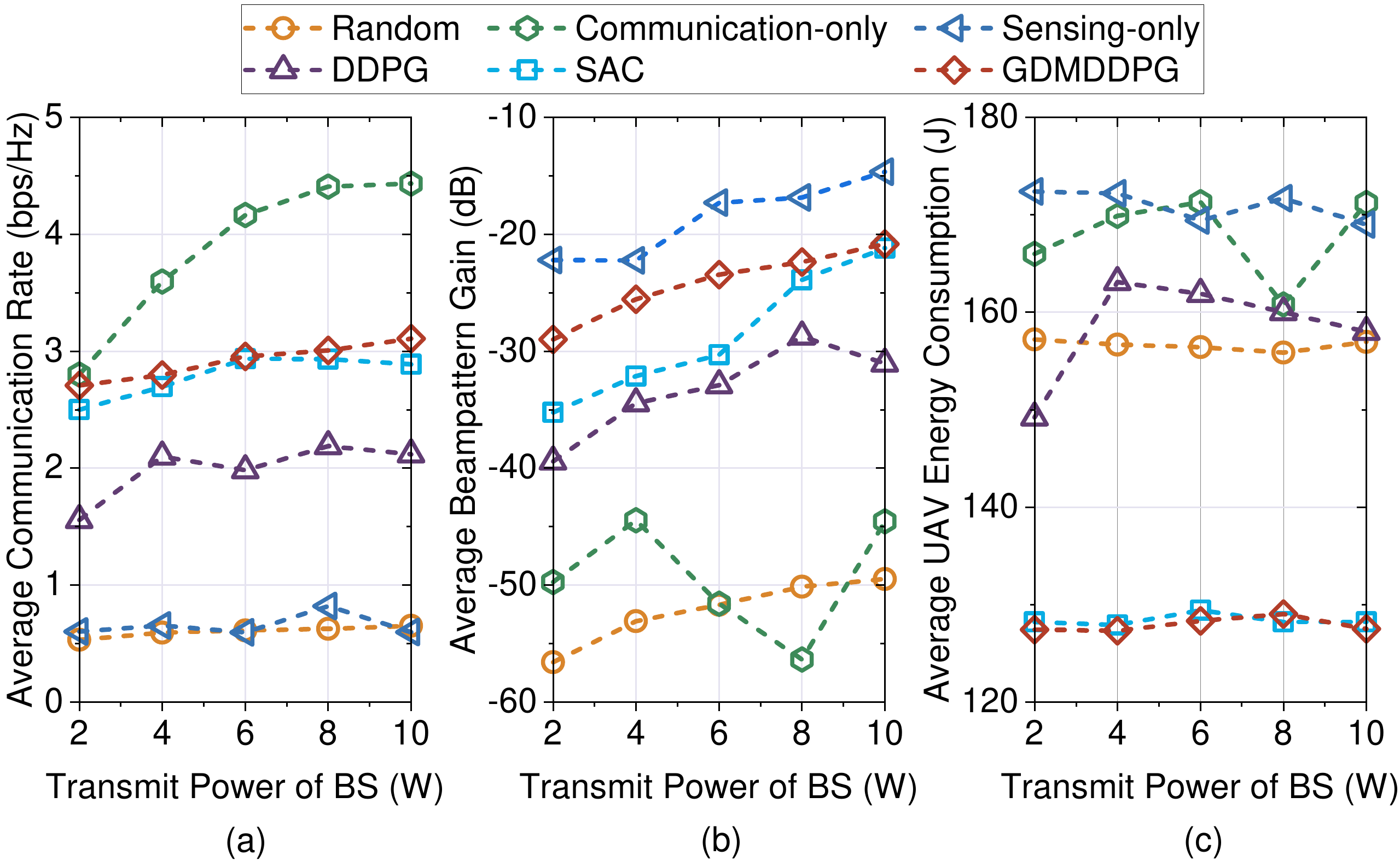}
    \caption{The impact of transmit power on different optimization objectives. (a) Average communication rate. (b) Average beampattern gain. (c) Average UAV energy consumption.}
    \label{fig:Power}
\end{figure}

\begin{figure}[t]
    \centering
    \includegraphics[width=\linewidth]{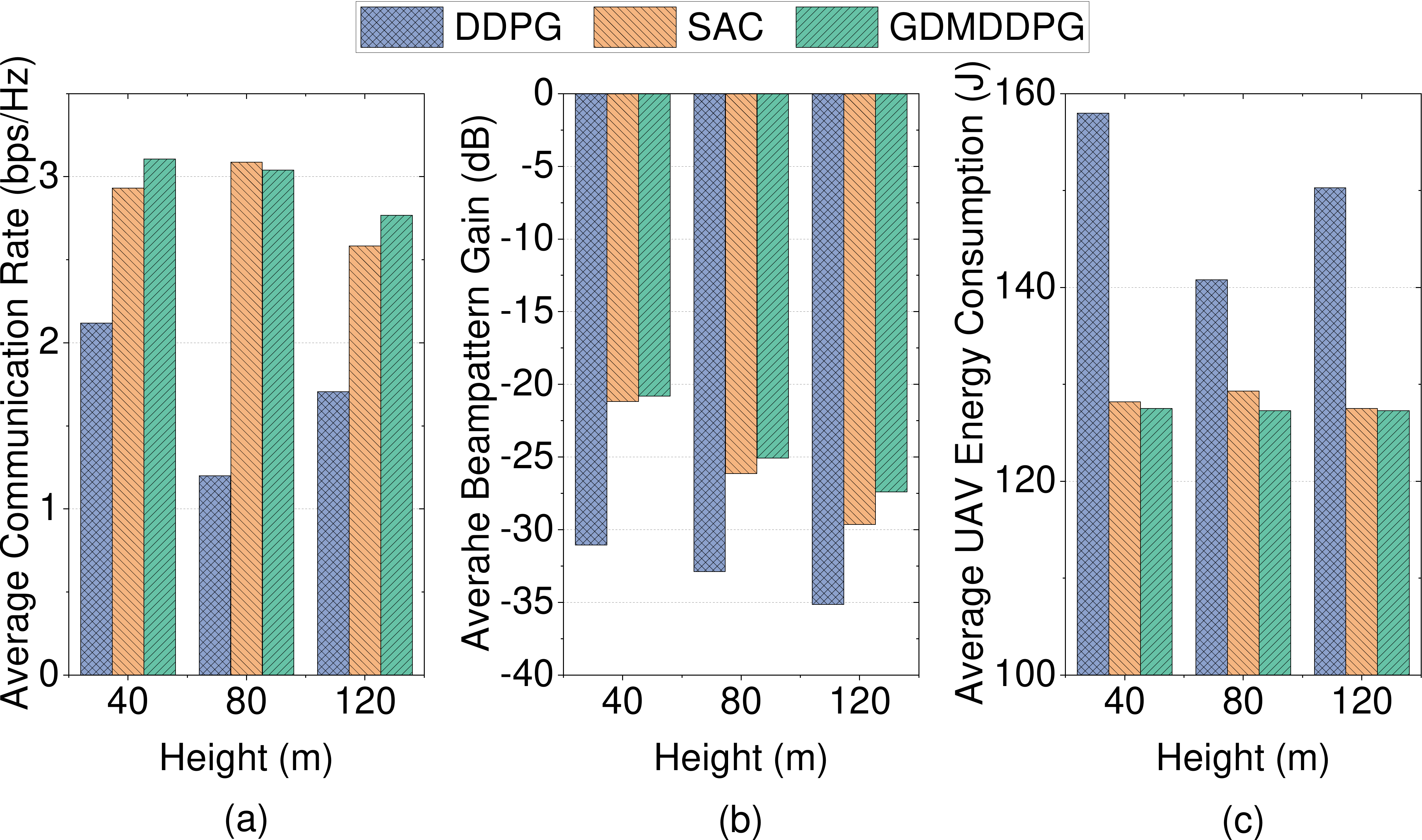}
    \caption{The impact of UAV flight height on different optimization objectives. (a) Average communication rate. (b) Average beampattern gain. (c) Average UAV energy consumption.}
    \label{fig:UAV Flight Height}
\end{figure}

\par Moreover, compared to both DDPG and SAC, the proposed GDMDDPG algorithm achieves superior performance across all three optimization metrics. This performance enhancement stems from three improvement factors proposed in this paper. First, the diffusion model strengthens the agent ability to analyze the environment state, thereby improving the quality of the decision-making process. Second, the noise perturbation mechanism enhances the exploratory capability of the agent within the action space. Finally, the RPER sampling mechanism, which integrates ERE and PER techniques, enables more efficient utilization of experience data. Specifically, it avoids forgetting past experiences while increasing the sampling frequency of recent experiences with high learning value. This experience sampling mechanism further boosts the learning efficiency of the algorithm. 

\subsection{Impact of System Settings}

\par In the following, we evaluate the performance of different algorithms under different system settings, including the transmit power of the BS, the UAV flight distance, and the number of BS antennas.
 
 \subsubsection{Impact of Maximum Transmit Power}
 \par Fig.~\ref{fig:Power} shows the trends of the three optimization objectives in the optimization problem with the variation of the BS maximum transmit power $P_{\rm max}$. The results demonstrate that as $P_{\rm max}$ increases, the communication rate and beampattern gain of DDPG, SAC, and the proposed GDMDDPG algorithms all show an upward trend. This occurs because the increasing $P_{\rm max}$ strengthens the signal, leading to improved communication and sensing capabilities. Moreover, the GDMDDPG algorithm achieves the lowest UAV energy consumption, demonstrating its effectiveness in trajectory design. In addition, the beampttern gain is extremely low in the communication-only method. This phenomenon occurs because the low-altitude ISR is deployed closer to the user cluster area and the passive beamforming of the IRS is fully aligned with the user, resulting in a significant attenuation of the sensing performance. In contrast, the user communication rate in the sensing-only method does not approach zero, likely due to the direct communication link between the BS and the users, which maintains a certain level of communication capability even under the sensing-priority condition. Furthermore, the random method performs poorly in all performance metrics due to the lack of systematic and theoretical basis for setting optimization variables by using a random strategy.

\begin{figure*}
    \centering
    \includegraphics[width=0.95\linewidth]{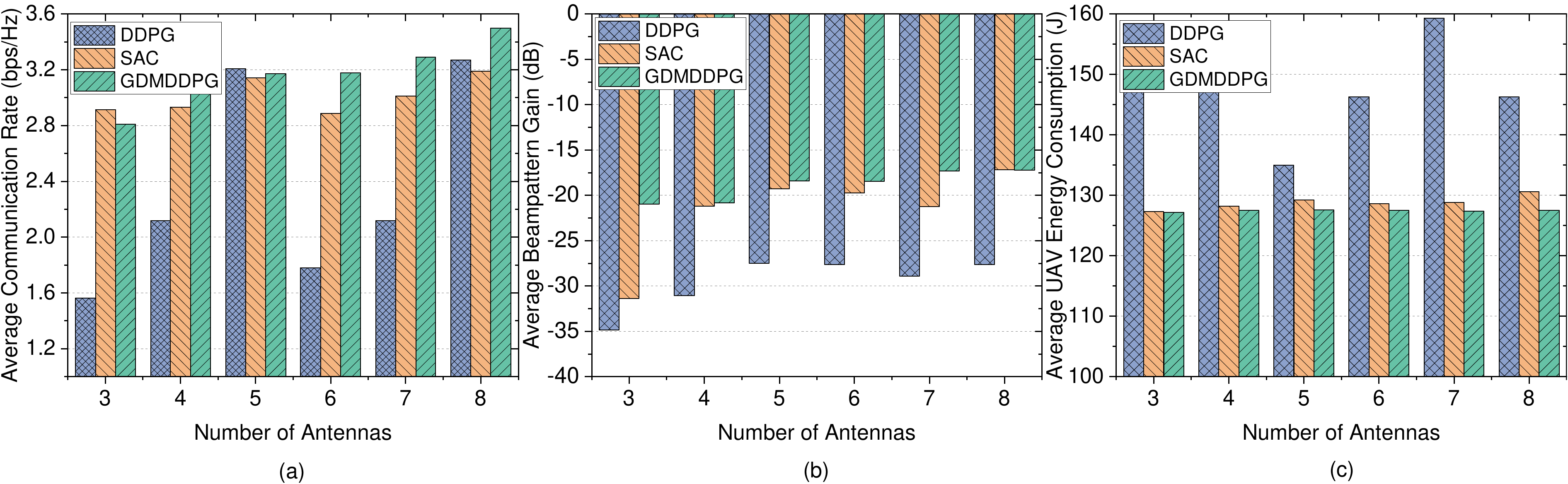}
    \caption{The impact of antenna number on different optimization objectives. (a) Average communication rate. (b) Average beampattern gain. (c) Average UAV energy consumption.}
    \label{fig:antenna}
\end{figure*}

\begin{figure*}[t]
    \centering
    \includegraphics[width=0.95\linewidth]{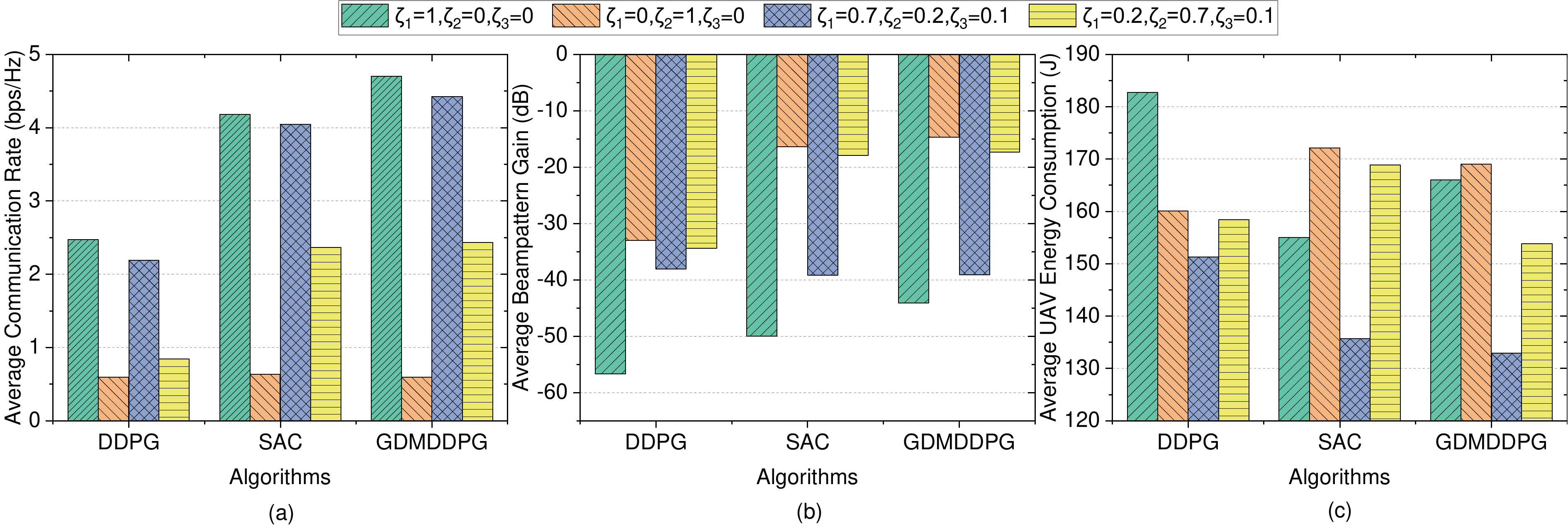}
    \caption{The impact of different importance weights on different optimization objectives. (a) Average communication rate. (b) Average beampattern gain. (c) Average UAV energy consumption.}
    \label{fig:Weights}
\end{figure*}

\subsubsection{Impact of UAV Flight Height} 
Fig.~\ref{fig:UAV Flight Height} shows the average communication rate, beampattern gain, and UAV energy consumption under different UAV flight heights. As can be seen, the average communication rate and beampattern gain show a downtrend as the UAV flight height increases. This phenomenon occurs because the higher UAV flight height leads to higher path loss, which reduces channel gain and consequently impacts communication and sensing performance. Moreover, we find that the beampattern gain decreases more significantly than the communication rate. This is primarily because direct communication links between the BS and users remain relatively unaffected by changes in UAV flight height. In addition, it can be observed that the proposed GDMDDPG algorithm outperforms the conventional DDPG algorithm and SAC algorithm regarding average communication rate, beampattern gain, and UAV energy consumption. The results indicate that the proposed algorithm remains effective across different UAV flight heights, mainly because the integrated diffusion model enhances the agent ability to understand environmental states.

\subsubsection{Impact of Antenna Number} 
\par Fig.~\ref{fig:antenna} shows the optimization objective values under different numbers of antennas. With more antennas, the communication rate and beampattern gain achieved by the GDMDDPG algorithm show an upward trend, as a larger array allows for more accurate beamforming and stronger energy focusing toward desired directions. Moreover, it can be observed that the performance of DDPG and SAC is unstable and even exhibits a degradation trend as the number of antennas increases. This phenomenon occurs because an increasing number of antennas enlarges the dimension of the action space, which makes it more difficult for DDPG and SAC to identify better solutions. In contrast, the above results indicate that the proposed GDMDDPG algorithm can still maintain superior exploration capability as the dimension of the action space increases.

\subsection{Optimization Results under Different Priorities}
In this part, we extend the reward function in Eq.~(\ref{eq:redesigned reward function}) to represent the optimization objectives with different priorities by introducing the importance weights. Specifically, the reward function with importance weights can be expressed as $r[t] = \zeta_{1}\eta_{1}R^{\text{U}}[t] + \zeta_{2}\eta_{2} G'[t] + \zeta_{3}\eta_{3}E'[t]-PV$, where $\zeta_1 \in [0,1]$, $\zeta_2 \in [0,1]$, and $\zeta_3 \in [0,1]$ denote the importance weights with $\zeta_1+\zeta_2+\zeta_3=1$. In practical ISAC applications, the priorities among sensing, communication, and energy consumption are typically determined by network operators based on specific mission priorities. In this case, the robustness of the proposed algorithm to different weight configurations is crucial.

\begin{figure}[t]
    \centering
    \includegraphics[width=0.71\linewidth]{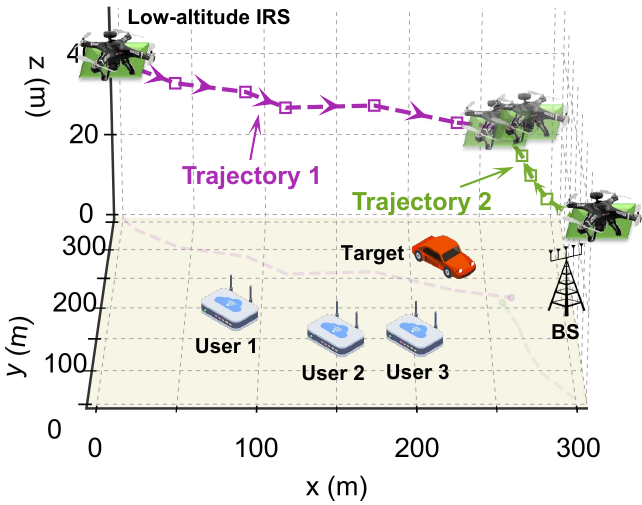}
    \caption{The trajectories of the UAV. With the same spatial distribution of BS, users, and target, the UAVs start from different initial locations.}
    \label{fig:Trajectory}
\end{figure}

\par Fig.~\ref{fig:Weights} shows the optimization values obtained by different algorithms under various importance weight configurations. As we can see, as the importance weights shift from favoring the communication objective, to favoring the sensing objective, to considering multiple objectives, the solutions obtained by the algorithms exhibit completely different trade-off patterns, which indicates that we can flexibly adjust the importance weights to accommodate the requirements of diverse application scenarios. Moreover, the GDMDDPG algorithm demonstrates superior overall performance across all importance weight settings compared with other comparison algorithms. Specifically, when only communication performance or sensing performance is considered, the GDMDDPG algorithm achieves the best communication rate and beampattern gain, respectively. In addition, when all three objectives are assigned non-zero weights, the GDMDDPG algorithm is able to prioritize the preferred objective while still improving the other two, which leads to a significantly better overall performance than the comparison algorithms. As such, the simulation results confirm that the proposed algorithm can efficiently solve the complex multi-objective optimization problem under different mission priority settings while maintaining stable and competitive performance.

\subsection{UAV Trajectory Analysis}
\par The schematic of the UAV trajectory under the spatial distribution of the BS, users, and target is shown in Fig.~\ref{fig:Trajectory}. Specifically, the initial location of the UAV is $(0 \ \text{m}, 300 \ \text{m}, 40 \ \text{m})$ in the first simulation. As can be seen, the UAV flies toward the clustered region of the users, target, and BS to efficiently perform communication and sensing tasks, and then hovers at its final location. Moreover, to validate the robustness of the proposed algorithm, repeated simulations are carried out with the initial location of the UAV set to $(300 
\ \text{m}, 0 \ \text{m}, 40 \ \text{m})$. The final hovering location in the second simulation is similar to that of the first simulation, further confirming the consistency and stability of the GDMDDPG algorithm.

%
\section{Conclusion}
\label{sec: Conclusion}
\par In this paper, we have investigated a low-altitude IRS-assisted ISAC system, where a BS simultaneously provides communication and sensing services to multiple users and a target. Moreover, we have formulated an optimization problem to maximize the communication rate and beampattern gain, while minimizing the UAV propulsion energy consumption by jointly optimizing the BS beamforming matrix, IRS phase shifts, UAV flight velocity, and UAV flight angle. Moreover, we have proposed the GDMDDPG algorithm, which improves the decision quality and learning efficiency by integrating the diffusion model, noise perturbation mechanism, and RPER sampling mechanism. Simulation results confirm that the proposed GDMDDPG algorithm can effectively achieve the simultaneous optimization of communication, sensing, and energy efficiency, thereby outperforming the other benchmark methods. In addition, GDMDDPG exhibits significant advantages in terms of robustness.

\bibliography{main}

@Article{Qaisar2026,
  author    = {Muhammad Umar Farooq Qaisar and Weijie Yuan and Onur G{\"{u}}nl{\"{u}} and Taneli Riihonen and Yuanhao Cui and Lin Zhang and Nuria {Gonz{\'{a}}lez Prelcic} and Marco Di Renzo and Zhu Han},
  journal   = {{IEEE} Trans. Netw. Sci. Eng.},
  title     = {The Role of {ISAC} in 6{G} Networks: Enabling Next-Generation Wireless Systems},
  year      = {2026},
  pages     = {7825--7861},
  volume    = {13},
}

@Article{Yang2026,
  author    = {Mengmeng Yang and Youyang Qu and Thilina Ranbaduge and Chandra Thapa and Nazatul Haque Sultan and Ming Ding and Hajime Suzuki and Wei Ni and Sharif Abuadbba and David B. Smith and Paul Tyler and Josef Pieprzyk and Thierry Rakotoarivelo and Xinlong Guan and Sirine Mrabet},
  journal   = {{ACM} Comput. Surv.},
  title     = {From 5{G} to 6{G}: {A} Survey on Security, Privacy, and Standardization Pathways},
  year      = {2026},
  number    = {8},
  pages     = {193:1--193:38},
  volume    = {58},
}

@Article{Chen2026,
  author    = {Guangji Chen and Qingqing Wu and Shihang Lu and Meng Hua and Wen Chen},
  journal   = {{IEEE} Trans. Wirel. Commun.},
  title     = {Multi-{IRS}-Aided {ISAC} System: Multi-Path Exploitation Versus Reduction},
  year      = {2026},
  pages     = {13351--13368},
  volume    = {25},
}

@Article{Ning2026,
  author    = {Zhaolong Ning and Yuzhen Zhang and Xiaojie Wang and Lei Guo and Dusit Niyato and Yan Zhang},
  journal   = {{IEEE} Trans. Wirel. Commun.},
  title     = {Joint Trajectory and Beamforming Optimization for {UAV-ISAC} Secure Communications},
  year      = {2026},
  pages     = {15216--15231},
  volume    = {25},
}

@ARTICLE{11543384,
  author={Liu, Wenchao and Zhang, Xuhui and Ren, Jinke and Yuan, Weijie and You, Changsheng and Li, Shuangyang},
  journal={{IEEE} Trans. Commun.}, 
  title={{UAV}-Enabled {ISAC} With Fluid Antennas for Low-Altitude Wireless Networks}, 
  year={2026},
  volume={74},
  number={},
  pages={9529-9546},
}

@ARTICLE{11395974,
  author={Zhang, Jieling and Tang, Huijun and Jiao, Pengfei and Wu, Huaming and Zhao, Zhidong and Li, Ruidong},
  journal={{IEEE} Internet Things J.}, 
  title={Secrecy Rate Optimization Based on {GNN} for {RIS}-Assisted {ISAC} System}, 
  year={2026},
  volume={13},
  number={12},
  pages={28065-28068},
}

@Article{11177504,
  author  = {Cao, Xiaowen and Jiang, Peng and Zhu, Guangxu and He, Yejun and Guizani, Moshen},
  journal = {{IEEE} Trans. Netw. Sci. Eng.},
  title   = {Joint Antenna Position and Beamforming Optimization for Movable Antenna Enabled Secure {IRS}-{ISAC} Network},
  year    = {2025},
  pages   = {1-15},
}

@InProceedings{Colas2018,
  author    = {C{\'{e}}dric Colas and Olivier Sigaud and Pierre{-}Yves Oudeyer},
  booktitle = {Proceedings of the 35th International Conference on Machine Learning, {ICML} 2018},
  title     = {{GEP-PG:} Decoupling Exploration and Exploitation in Deep Reinforcement Learning Algorithms},
  year      = {2018},
  pages     = {1038--1047},
  volume    = {80},
}

@Article{Hua2023,
  author    = {Haocheng Hua and Jie Xu and Tony Xiao Han},
  journal   = {{IEEE} Trans. Veh. Technol.},
  title     = {Optimal Transmit Beamforming for Integrated Sensing and Communication},
  year      = {2023},
  number    = {8},
  pages     = {10588--10603},
  volume    = {72},
}

@Article{Lyu2025,
  author    = {Wanting Lyu and Songjie Yang and Yue Xiu and Zhongpei Zhang and Chadi Assi and Chau Yuen},
  journal   = {{IEEE} Trans. Wirel. Commun.},
  title     = {Movable Antenna Enabled Integrated Sensing and Communication},
  year      = {2025},
  number    = {4},
  pages     = {2862--2875},
  volume    = {24},
}

@Article{11159297,
  author  = {Wu, Jun and Yuan, Weijie and Cheng, Qingqing and Jin, Haijia},
  journal = {{IEEE} J. Sel. Areas Commun.},
  title   = {Toward Dual-Functional {LAWN}: Control-Aware System Design for Aerodynamics-Aided {UAV} Formations},
  year    = {2025},
  pages   = {1-1},
}

@Article{11165100,
  author  = {Yang, Ruihang and Wang, Dezhi and Zhu, Chen and Ning, Boyu and Zhu, Zhengyu and Huang, Chongwen and Yang, Zhaohui},
  journal = {{IEEE} Trans. Veh. Technol.},
  title   = {Beamforming Design for {RIS}-aided Integrated Sensing, Communication, and Computation Systems},
  year    = {2025},
  pages   = {1-14},
}

@Article{Wu2023,
  author    = {Jun Wu and Weijie Yuan and Lin Bai},
  journal   = {{IEEE} Internet Things J.},
  title     = {On the Interplay Between Sensing and Communications for {UAV} Trajectory Design},
  year      = {2023},
  number    = {23},
  pages     = {20383--20395},
  volume    = {10},
}

@Article{Jin2025,
  author    = {Haijia Jin and Jun Wu and Weijie Yuan and Fan Liu and Yuanhao Cui},
  journal   = {{IEEE} Trans. Mob. Comput.},
  title     = {Co-Design of Sensing, Communications, and Control for Low-Altitude Wireless Networks},
  year      = {2025},
  number    = {11},
  pages     = {12035--12048},
  volume    = {24},
}

@Article{Goek2024,
  author    = {Mehmet G{\"{o}}k},
  journal   = {Appl. Soft Comput.},
  title     = {Dynamic path planning via Dueling Double Deep {Q}-Network {(D3QN)} with prioritized experience replay},
  year      = {2024},
  pages     = {111503},
  volume    = {158},
}

@InProceedings{Haarnoja2018,
  author    = {Tuomas Haarnoja and Aurick Zhou and Pieter Abbeel and Sergey Levine},
  booktitle = {Proceedings of the 35th International Conference on Machine Learning, {ICML}},
  title     = {Soft Actor-Critic: Off-Policy Maximum Entropy Deep Reinforcement Learning with a Stochastic Actor},
  year      = {2018},
  pages     = {1856--1865},
  volume    = {80},
}

@Article{Wu2021tutorial,
  author    = {Qingqing Wu and Shuowen Zhang and Beixiong Zheng and others},
  journal   = {{IEEE} Trans. Commun.},
  title     = {Intelligent Reflecting Surface-Aided Wireless Communications: {A} Tutorial},
  year      = {2021},
  number    = {5},
  pages     = {3313--3351},
  volume    = {69},
}

@Article{Xu2024,
  author  = {Jinlei Xu and Dongdong Li and Zhengyu Zhu and Zhutian Yang and Nan Zhao and Dusit Niyato},
  journal = {{IEEE} Trans. Commun.},
  title   = {Anti-Jamming Design for Integrated Sensing and Communication via Aerial {IRS}},
  year    = {2024},
  number  = {8},
  pages   = {4607--4619},
  volume  = {72},
}

@Article{Zuo2023,
  author  = {Jiakuo Zuo and Yuanwei Liu and Chenming Zhu and Yixuan Zou and Dengyin Zhang and Naofal Al{-}Dhahir},
  journal = {{IEEE} Trans. Veh. Technol.},
  title   = {Exploiting {NOMA} and {RIS} in Integrated Sensing and Communication},
  year    = {2023},
  number  = {10},
  pages   = {12941--12955},
  volume  = {72},
}

@Article{Singh2025,
  author    = {Jitendra Singh and Awadhesh Gupta and Aditya K. Jagannatham and Lajos Hanzo},
  journal   = {{IEEE} Trans. Veh. Technol.},
  title     = {Multi-Beam Object-Localization for Millimeter-Wave {ISAC}-Aided Connected Autonomous Vehicles},
  year      = {2025},
  number    = {1},
  pages     = {1725--1729},
  volume    = {74},
}

@Article{Liao2023,
  author  = {Chikun Liao and Feng Wang and Vincent K. N. Lau},
  journal = {{IEEE} Trans. Commun.},
  title   = {Optimized Design for {IRS}-Assisted Integrated Sensing and Communication Systems in Clutter Environments},
  year    = {2023},
  number  = {8},
  pages   = {4721--4734},
  volume  = {71},
}

@Article{AlHourani2014,
  author  = {Akram Al{-}Hourani and Kandeepan Sithamparanathan and Simon Lardner},
  journal = {{IEEE} Wirel. Commun. Lett.},
  title   = {Optimal {LAP} Altitude for Maximum Coverage},
  year    = {2014},
  number  = {6},
  pages   = {569--572},
  volume  = {3},
}

@Article{Hu2023,
  author  = {Xiaoling Hu and Chenxi Liu and Mugen Peng and Caijun Zhong},
  journal = {{IEEE} Trans. Wirel. Commun.},
  title   = {{IRS}-Based Integrated Location Sensing and Communication for mm{W}ave {SIMO} Systems},
  year    = {2023},
  number  = {6},
  pages   = {4132--4145},
  volume  = {22},
}

@Article{Li2023,
  author  = {Jin Li and Gui Zhou and Tantao Gong and Nan Liu},
  journal = {{IEEE} Wirel. Commun. Lett.},
  title   = {Beamforming Design for Active {IRS}-Aided {MIMO} Integrated Sensing and Communication Systems},
  year    = {2023},
  number  = {10},
  pages   = {1786--1790},
  volume  = {12},
}

@Article{Fang2024,
  author  = {Yuan Fang and Siyao Zhang and Xinmin Li and Xianghao Yu and Jie Xu and Shuguang Cui},
  journal = {{IEEE} Trans. Commun.},
  title   = {Multi-{IRS}-Enabled Integrated Sensing and Communications},
  year    = {2024},
  number  = {9},
  pages   = {5853--5867},
  volume  = {72},
}

@Article{Hua2024,
  author  = {Meng Hua and Qingqing Wu and Wen Chen and Octavia A. Dobre and A. Lee Swindlehurst},
  journal = {{IEEE} Trans. Wirel. Commun.},
  title   = {Secure Intelligent Reflecting Surface-Aided Integrated Sensing and Communication},
  year    = {2024},
  number  = {1},
  pages   = {575--591},
  volume  = {23},
}

@Article{Wang2023,
  author  = {Xinyi Wang and Zesong Fei and Qingqing Wu},
  journal = {{IEEE} Internet Things J.},
  title   = {Integrated Sensing and Communication for {RIS}-Assisted Backscatter Systems},
  year    = {2023},
  number  = {15},
  pages   = {13716--13726},
  volume  = {10},
}

@Article{Jiang2024,
  author  = {Chengjun Jiang and Chensi Zhang and Chongwen Huang and Jianhua Ge and M{\'{e}}rouane Debbah and Chau Yuen},
  journal = {{IEEE} Internet Things J.},
  title   = {Exploiting {RIS} in Secure Beamforming Design for {NOMA}-Assisted Integrated Sensing and Communication},
  year    = {2024},
  number  = {17},
  pages   = {28123--28136},
  volume  = {11},
}

@Article{Long2024,
  author  = {Xudong Long and Yubin Zhao and Huaming Wu and Cheng{-}Zhong Xu},
  journal = {{IEEE} Internet Things J.},
  title   = {Deep Reinforcement Learning for Integrated Sensing and Communication in {RIS}-Assisted 6{G} {V2X} System},
  year    = {2024},
  number  = {24},
  pages   = {39834--39849},
  volume  = {11},
}

@Article{Hu2024,
  author  = {Langtao Hu and Rui Yang and Lei Wu and Chongwen Huang and Yu'e Jiang and Li Chen and Xiaobo Zhou},
  journal = {{IEEE} Internet Things J.},
  title   = {{RIS}-Assisted Integrated Sensing and Covert Communication Design},
  year    = {2024},
  number  = {9},
  pages   = {16505--16516},
  volume  = {11},
}

@Article{Cao2023,
  author  = {Xueyan Cao and Xiaoling Hu and Mugen Peng},
  journal = {{IEEE} Trans. Veh. Technol.},
  title   = {Feedback-Based Beam Training for Intelligent Reflecting Surface Aided mmWave Integrated Sensing and Communication},
  year    = {2023},
  number  = {6},
  pages   = {7584--7596},
  volume  = {72},
}

@InProceedings{10594249,
  author    = {Chen, Xipeng and Cao, Xiaowen and Xie, Lifeng and He, Yejun},
  booktitle = {IEEE International Workshop on Radio Frequency and Antenna Technologies (iWRF\&AT)},
  title     = {{DRL}-Based Joint Trajectory Planning and Beamforming Optimization in Aerial {RIS}-Assisted {ISAC} System},
  year      = {2024},
  pages     = {510-515},
}

@InProceedings{10257639,
  author    = {Zhu, Zhengyu and Gong, Mengfei and Chu, Zheng and Xiao, Pei and Sun, Gangcan and Mi, De and He, Ziming and Tong, Fei},
  booktitle = {International Conference on Ubiquitous Communication (Ucom)},
  title     = {{DRL}-based {STAR}-{RIS}-Assisted {ISAC} Secure Communications},
  year      = {2023},
  pages     = {127-132},
}

@Article{Du2024a,
  author  = {Hongyang Du and Zonghang Li and Dusit Niyato and Jiawen Kang and Zehui Xiong and Huawei Huang and Shiwen Mao},
  journal = {{IEEE} Trans. Mob. Comput.},
  title   = {Diffusion-Based Reinforcement Learning for Edge-Enabled {AI}-Generated Content Services},
  year    = {2024},
  number  = {9},
  pages   = {8902--8918},
  volume  = {23},
}

@Article{Yang2024Diffusion,
  author    = {Ling Yang and Zhilong Zhang and Yang Song and Shenda Hong and Runsheng Xu and Yue Zhao and Wentao Zhang and Bin Cui and Ming{-}Hsuan Yang},
  journal   = {{ACM} Comput. Surv.},
  title     = {Diffusion Models: {A} Comprehensive Survey of Methods and Applications},
  year      = {2024},
  number    = {4},
  pages     = {105:1--105:39},
  volume    = {56},
}

@Article{Wang2019,
  author        = {Che Wang and Keith W. Ross},
  journal       = {CoRR},
  title         = {Boosting Soft Actor-Critic: Emphasizing Recent Experience without Forgetting the Past},
  year          = {2019},
  volume        = {abs/1906.04009},
  archiveprefix = {arXiv},
}

@InProceedings{Lillicrap2016,
  author    = {Timothy P. Lillicrap and Jonathan J. Hunt and Alexander Pritzel and Nicolas Heess and Tom Erez and Yuval Tassa and David Silver and Daan Wierstra},
  booktitle = {4th International Conference on Learning Representations, {ICLR}},
  title     = {Continuous control with deep reinforcement learning},
  year      = {2016},
}

@Article{10759668,
  author  = {Li, Zhuoran and Gao, Zhen and Wang, Kuiyu and Mei, Yikun and Zhu, Chunli and Chen, Lei and Wu, Xiaomei and Niyato, Dusit},
  journal = {{IEEE} Internet Things J.},
  title   = {Unauthorized {UAV} Countermeasure for Low-Altitude Economy: Joint Communications and Jamming Based on {MIMO} Cellular Systems},
  year    = {2024},
  pages   = {1-1},
}

@Article{Kaushik2024,
  author  = {Aryan Kaushik and Rohit Singh and Shalanika Dayarathna and Rajitha Senanayake and Marco Di Renzo and Miguel Dajer and Hyoungju Ji and Younsun Kim and Vincenzo Sciancalepore and Alessio Zappone and Wonjae Shin},
  journal = {{IEEE} Commun. Stand. Mag.},
  title   = {Toward Integrated Sensing and Communications for 6{G}: Key Enabling Technologies, Standardization, and Challenges},
  year    = {2024},
  number  = {2},
  pages   = {52--59},
  volume  = {8},
}

@Article{Geng2025,
  author    = {Yue Geng and Tee Hiang Cheng and Kai Zhong and Kah Chan Teh and Qingqing Wu},
  journal   = {{IEEE} Trans. Wirel. Commun.},
  title     = {Joint Beamforming for {CRB}-Constrained {IRS}-Aided {ISAC} System via Product Manifold Methods},
  year      = {2025},
  number    = {1},
  pages     = {691--705},
  volume    = {24},
}

@Article{He2022,
  author    = {Yinghui He and Yunlong Cai and Hao Mao and Guanding Yu},
  journal   = {{IEEE} J. Sel. Areas Commun.},
  title     = {{RIS}-Assisted Communication Radar Coexistence: Joint Beamforming Design and Analysis},
  year      = {2022},
  number    = {7},
  pages     = {2131--2145},
  volume    = {40},
}

@Article{Sun2025-1,
  author    = {Geng Sun and Wenwen Xie and Dusit Niyato and Fang Mei and Jiawen Kang and Hongyang Du and Shiwen Mao},
  journal   = {{IEEE} Wirel. Commun.},
  title     = {Generative {AI} for Deep Reinforcement Learning: Framework, Analysis, and Use Cases},
  year      = {2025},
  number    = {3},
  pages     = {186--195},
  volume    = {32},
}

@Article{Sun2025a,
  author    = {Geng Sun and Jian Xiao and Jiahui Li and Jiacheng Wang and Jiawen Kang and Dusit Niyato and Shiwen Mao},
  journal   = {{IEEE} Trans. Mob. Comput.},
  title     = {Aerial Reliable Collaborative Communications for Terrestrial Mobile Users via Evolutionary Multi-Objective Deep Reinforcement Learning},
  year      = {2025},
  number    = {7},
  pages     = {5731--5748},
  volume    = {24},
}

@Article{Liu2022,
  author  = {Fan Liu and Yuanhao Cui and Christos Masouros and Jie Xu and Tony Xiao Han and Yonina C. Eldar and Stefano Buzzi},
  journal = {{IEEE} J. Sel. Areas Commun.},
  title   = {Integrated Sensing and Communications: Toward Dual-Functional Wireless Networks for 6{G} and Beyond},
  year    = {2022},
  number  = {6},
  pages   = {1728--1767},
  volume  = {40},
}

@Article{Sankar2024,
  author  = {R. S. Prasobh Sankar and Sundeep Prabhakar Chepuri and Yonina C. Eldar},
  journal = {{IEEE} Trans. Wirel. Commun.},
  title   = {Beamforming in Integrated Sensing and Communication Systems With Reconfigurable Intelligent Surfaces},
  year    = {2024},
  number  = {5},
  pages   = {4017--4031},
  volume  = {23},
}

@Article{Guo2023,
  author  = {Shaoai Guo and Xiaohui Zhao},
  journal = {{IEEE} Trans. Commun.},
  title   = {Multi-Agent Deep Reinforcement Learning Based Transmission Latency Minimization for Delay-Sensitive Cognitive Satellite-{UAV} Networks},
  year    = {2023},
  number  = {1},
  pages   = {131--144},
  volume  = {71},
}

@Article{Zhang2025a,
  author    = {Chuang Zhang and Geng Sun and Jiahui Li and Qingqing Wu and Jiacheng Wang and Dusit Niyato and Yuanwei Liu},
  journal   = {{IEEE} Trans. Mob. Comput.},
  title     = {Multi-Objective Aerial Collaborative Secure Communication Optimization via Generative Diffusion Model-Enabled Deep Reinforcement Learning},
  year      = {2025},
  number    = {4},
  pages     = {3041--3058},
  volume    = {24},
}

@Article{Oubbati2022,
  author    = {Omar Sami Oubbati and Mohammed Atiquzzaman and Hyotaek Lim and Abderrezak Rachedi and Abderrahmane Lakas},
  journal   = {{IEEE} Trans. Veh. Technol.},
  title     = {Synchronizing {UAV} Teams for Timely Data Collection and Energy Transfer by Deep Reinforcement Learning},
  year      = {2022},
  number    = {6},
  pages     = {6682--6697},
  volume    = {71},
}

@Article{Dai2022,
  author    = {Xinhong Dai and Bin Duo and Xiaojun Yuan and Wanbin Tang},
  journal   = {{IEEE} Wirel. Commun. Lett.},
  title     = {Energy-Efficient {UAV} Communications: {A} Generalized Propulsion Energy Consumption Model},
  year      = {2022},
  number    = {10},
  pages     = {2150--2154},
  volume    = {11},
}

@Article{Liang2024,
  author    = {Guangming Liang and Jie Hu and Yizhe Zhao and Kun Yang},
  journal   = {{IEEE} Trans. Commun.},
  title     = {Intelligent Link Adaptation for Integrated Data and Energy Transfer: An Enhanced {DRL} Approach for Long-Term Constraints},
  year      = {2024},
  number    = {11},
  pages     = {6956--6972},
  volume    = {72},
}

@Article{10839238,
  author  = {Sun, Geng and Xie, Wenwen and Niyato, Dusit and Mei, Fang and Kang, Jiawen and Du, Hongyang and Mao, Shiwen},
  journal = {{IEEE} Wireless Communications},
  title   = {Generative {AI} for Deep Reinforcement Learning: Framework, Analysis, and Use Cases},
  year    = {2025},
  pages   = {1-10},
}

@Article{Xie2026,
  author    = {Wenwen Xie and Geng Sun and Bei Liu and Jiahui Li and Jiacheng Wang and Hongyang Du and Dusit Niyato and Dong In Kim},
  journal   = {{IEEE} Trans. Mob. Comput.},
  title     = {Joint Optimization of {UAV}-Carried {IRS} for Urban Low Altitude mm{W}ave Communications With Deep Reinforcement Learning},
  year      = {2026},
  number    = {1},
  pages     = {1381--1397},
  volume    = {25},
}

\end{document}